\documentstyle[preprint,aps]{revtex}
\draft
\begin{document}
\bibliographystyle{prsty}
\preprint{\vbox{\hbox{RUB-TPII-7/95} \hbox{hep-ph/9504363}}}
\title{
Electromagnetic Form Factors of the SU(3) Octet Baryons     \\
in the semibosonized SU(3) Nambu-Jona-Lasinio Model}

\author{
Hyun-Chul Kim
\footnote{E-mail address: kim@hadron.tp2.ruhr-uni-bochum.de},
Andree Blotz
\footnote{Present address: Department of Physics, State University
of New York, Stony Brook, 11794, U.S.A.},
Maxim Polyakov
\footnote{On leave of absence from Petersburg Nuclear Physics
Institute, Gatchina, St. Petersburg 188350, Russia}
and Klaus Goeke
\footnote{E-mail address: goeke@hadron.tp2.ruhr-uni-bochum.de  }  }

\address{
Institute for  Theoretical  Physics  II, \\  P.O. Box 102148,
Ruhr-University Bochum, \\
D--44780 Bochum, Germany  \\
       }
\date{ \today  }
\maketitle
\begin{abstract}
The electromagnetic form factors of the SU(3) octet baryons
are investigated in the semibosonized SU(3) Nambu--Jona-Lasinio
model (chiral quark-soliton model).
The rotational $1/N_c$ and strange quark mass corrections
in linear order are taken into account.
The electromagnetic charge radii of the nucleon
and magnetic moments are also evaluated.
It turns out that the model is in a remarkable good agreement with
the experimental data.
\end{abstract}
\pacs{12.40.-y, 13.40.Em, 13.40.Gp, 14.20.Dh, 14.20.Jn}

\section{Introduction}
In spite of the belief that Quantum Chromodynamics (QCD)
is the fundamental underlying theory of the strong interaction,
low energy phenomena such as static properties of hadrons
defy solutions based on QCD.  The pertinacity of QCD in the
low energy region have led to efforts to construct an effective
theory for the strong interaction.
In pursuit of this aim, the chiral quark soliton
model--also known as the
semibosonized Nambu-Jona-Lasinio(NJL) model--emerged
as a successful effective theory
to describe the low energy phenomena without loss of
important properties of QCD such as chiral symmetry and
its spontaneous breaking.

Originally, the idea of finding the soliton
in a model with quarks coupled to pions was realized
by Kahana, Ripka and Soni~\cite{krs} and Banerjee and Birse~\cite{bb}.
The bound states of the valence quarks were well explored in the
model while it suffered from the vacuum instability~\cite{soni}.
This problem of the vacuum instability was solved
by Diakonov and Petrov~\cite{dp}.
Having investigated the instanton picture of the QCD vacuum
in the low-momenta limit in
Ref.~\cite{dp}, they have shown
that the low-momenta theory is equivalent to
the quark-soliton model free from the vacuum instability.
The model was further elaborated in Ref.\ \cite{dpp} so that it could
predict the static properties of the nucleon
in the gradient approximation.

The baryon in this model is regarded as $N_c$ valence
quarks coupled to the polarized
Dirac sea bound by a non-trivial chiral field configuration
in the Hartree approximation~\cite{dpp,rw,mgg,wy}.
The identification of the baryon quantum numbers
is acquired by the semi-classical
quantization~\cite{dpp,anw} (in nuclear physics
called the cranking method ~\cite{rs})
which is performed by integrating
over the zero-mode fluctuations of the pion field around the
saddle point.
It makes the baryon carry proper quantum numbers like spins
and isospins.  In SU(2), the model enables us to describe
quantitatively a great deal of static properties of the
nucleon such as $N$--$\Delta$ splitting~\cite{wy,getal},
axial constants~\cite{wy,ww,chetal1},
electromagnetic form factors~\cite{Wakamatsu,chetal},
and to some extent also magnetic moments~\cite{wy,chetal}.

Although the SU(2) version of the model was quite successful to
explain many static properties of the nucleon, it is necessary
to extend the model from SU(2) to SU(3) so that it can be possible to
examine the same properties of hyperons and moreover
to investigate the effects of hidden strangeness on the nucleon which
are in particular manifested in the $\pi N$ sigma term
\cite{blotz,bkg},
the iso--splitting of the baryonic masses~\cite{pbg1} and
strange form factors~\cite{bhg}.
Blotz {\em et al.}~\cite{betal1,betal2}
and Weigel {\em et al.}~\cite{war}
have carried out the extension of the
model from SU(2) to SU(3).  Starting from the semibosonized
NJL-type lagrangian, they have shown that
the model describes hyperon spectra successfully.
The extended SU(3) model is distinguished from the SU(2) NJL
in two ways: Firstly, the mixed terms of the pure SU(2) part and
and the strange vacuum part are induced by the trivial embedding
of the SU(2) soliton into SU(3).
Secondly, since the mass of the
strange is not negligible, one has to take into account
the mass term in the effective action explicitly.
The mass corrections are treated perturbatively in linear order.
It was shown that the perturbative treatment of the $m_s$
in the NJL model describes the octet-decuplet mass splitting
\cite{betal1,betal2} very well and
plays an essential role in the mass splitting of hyperons.
These two differences determine the characteristic of the SU(3)
NJL model.

Refs.~\cite{betal1,war} indicate that the SU(3) NJL provides
a more refined structure of the collective hamiltonian than the
pseudoscalar Skyrme model.  A comparable structure can be obtained
in the Skyrme model only by introducing explicit vector mesons.
However, it is inevitable to
import large numbers of parameters into the Skyrme model with
vector meson, while the parameters in the NJL model can be
fixed completely by adjusting mesonic masses and decay constants
($f_{\pi}$, $f_{K}$).  The only free parameter we have is the
constituent quark mass arising as a result of the spontaneously
broken chiral symmetry.  This parameter is fixed by adjusting
the mass splitting~\cite{betal2} properly.

It is of great importance that $1/N_c$ rotational corrections
are taken into account.
Starting from the path integral formalism, when we
integrate over zero modes fluctuations around the saddle point,
a time ordered product of collective operators appears.
The $1/N_c$ contribution
survives due to the noncommutivity of the collective operators
\cite{ww}.
It was examined in detail in ref.~\cite{chetal1} by
calculating the axial vector constants $g_A$ and isovector
magnetic moments in SU(2).  In the same spirit,
the SU(3) model was applied to obtain the axial
constants $g^{(3)}_A$, $g^{(8)}_A$, and
$g^{(0)}_A$~\cite{bpg2,bpg3,blpolg}.
It predicted the experimental data within about $10 \%$.

In recent papers, we have proceeded to evaluate the
magnetic moments~\cite{kbpg}.  The magnetic moments of
the SU(3) octet baryons predicted by the present model
are in a remarkable agreement with the experiments.

Now, we are in a position to study the electromagnetic form factors
and other form factors such as strange form factors.
It is important to investigate the form factors in our model, since
it allows us to take a step forward in studying dynamics.
Hence, as a first phase, we will consider the electromagnetic
form factors.  It is of great significance to know them in the
SU(3) NJL in that not only they provide
us with the electromagnetic informations but also they
allow us to proceed to explore the techniques
for the form factors of the
neutral($Z^0$) currents and charged weak($W^{\pm}$) currents.

The outline of the paper is as follows.  In the next section, we develop
the general formalism for the electromagnetic form factors in the SU(3)
NJL.  In section 3, we discuss the electric form factors
with related quantities such as electric charge radii.  In section 4,
we continue to study the magnetic form factors of the SU(3) octet baryons.
In section 5, we summarize the work and draw conclusions.

\section{The General formalism}
In this section, we present the general formalism for the
electromagnetic form factors of the SU(3) octet baryons in the
NJL.

The SU(3) NJL is characterized by
a partition function in Euclidean space
given by the functional integral over pseudoscalar meson
and quark fields:
\begin{eqnarray}
{\cal Z} & = & \int {\cal D}\Psi {\cal D}\Psi^\dagger
{\cal D}\pi^a \exp{(-S_{NJL})} \nonumber \\
& = & \int {\cal D}\Psi {\cal D}\Psi^\dagger
{\cal D}\pi^a \exp\left (-\int d^4 x \Psi^\dagger i D \Psi \right ) ,
\label{eq:Z}
\end{eqnarray}
where $D$ denotes the Dirac differential operator
\begin{equation}
i D \; = \; \beta (- i \rlap{/}{\partial} + \hat{m} + MU)
\end{equation}
with the pseudoscalar chiral field
\begin{equation}
U\;=\;\exp{i\pi^a \lambda^a \gamma_{5}} .
\end{equation}
$\hat{m}$ is the matrix of the current quark mass given by
\begin{equation}
\hat{m}\;=\;diag(m_u, m_d, m_s) = m_0 {\mbox {\bf 1}} \;+\; m_8 \lambda_8.
\label{Eq:mass}
\end{equation}
$\lambda^a$ represent the usual Gell-Mann matrices normalized as
${\rm tr}\ {(\lambda^a \lambda^b)}=2\delta^{ab}$.  Here, we assume
isospin symmetry, {\em i.e.} $m_u=m_d$.
 $M$ shows the dynamical quark mass arising from the spontaneous
chiral symmetry breaking, which is in general momentum--dependent
\cite{dp}.
 For the sake of convenience we shall look upon $M$ as a constant and
introduce the ultra--violet cut--off via the proper time
regularization which preserves gauge and chiral invariance~\cite{ball}.
The $m_0$ and $m_8$ in eq.~(\ref{Eq:mass}) are respectively defined
by
\begin{equation}
m_0\;=\; \frac{m_u+m_d+m_s}{3},\;\;\;\;\;m_8\;=\;
\frac{m_u+m_d-2m_s}{2\sqrt{3}}.
\end{equation}
The operator $i D$ is expressed in Euclidean space in terms of
the Euclidean time derivative $\partial_\tau$
and the Dirac one--particle hamiltonian $H(U)$
\begin{equation}
i D \; = \; \partial_\tau \; + \; H(U)\; +\; \beta\hat{m}
\;-\; \beta\bar{m}{\bf 1}
\label{Eq:Dirac}
\end{equation}
with
\begin{equation}
H(U) \; = \; \frac{\vec{\alpha}\cdot \nabla}{i}
\;+\; \beta MU \;+\; \beta \bar{m} {\bf 1}.
\label{Eq:hamil}
\end{equation}
$\beta$ and $\vec{\alpha}$ are the well--known
Dirac hermitian matrices~\cite{iz}.  The $\bar{m}$ is defined by
$(m_u+m_d)/2 = m_u = m_d$.  We want to emphasize that the NJL model
is a low-energy effective model of QCD.  Hence, the action
of this model can have, in principle, corrections from higher orders
such as a term $\sim \;m^2 \Psi^\dagger \Psi$ for example.  However,
the coefficient in front of such a term is not known
\footnote{The coefficient for the $\hat{m}\Psi^\dagger \Psi$
is determined by the soft-pion theorem.} theoretically.
Therefore, it is meaningless
to go beyond the linear order of the quark mass expansion
unless higher order corrections ({\em e.g.} the coefficient in front
of $\hat{m}^2 \Psi^\dagger \Psi$) to the action are known.

The electromagnetic form factors of the baryons $F_i (q^2)$
are defined by the expectation values of the electromagnetic current
$V_\mu$ of the quark fields:
\begin{equation}
\langle B', p'| V_\mu(0) |B, p \rangle \;=\; \bar{u}_{B'} (p')
\left[ \gamma_\mu F_1 (q^2) \;+\;
i \sigma_{\mu\nu} \frac{q^\nu}{2M_N} F_2 (q^2)
\right ] u_{B}(p)
\end{equation}
with
\begin{equation}
V_\mu(z) \;=\; \bar{\Psi} (z) \gamma_\mu \hat{Q} \Psi(z).
\end{equation}
$M_N$ denotes the nucleon mass.
 $\hat{Q}$ designates the charge operator of the quark field $\Psi (z)$
\begin{eqnarray}
\hat{Q} &  = &  \left ( \begin{array}{ccc}
\frac{2}{3} & 0 & 0\\
0 & -\frac{1}{3} & 0 \\
0 & 0 & -\frac{1}{3} \end{array} \right ) \nonumber \\
& = & T_3 \; + \; \frac{Y}{2}.
\end{eqnarray}
$T_3$ and $Y$ are respectively the third component of the isospin
and hypercharge given by the Gell-Mann--Nishjima formula.
 The $q^2$ is just the four momentum transfer
$q^2 = -Q^2$ with $Q^2 >0$.
 Hence, the electromagnetic current $V_\mu$
can be decomposed into the third and eighth $SU(3)$
octet currents
\begin{equation}
V_\mu \;=\; V^{(3)}_\mu \; + \; \frac{1}{\sqrt{3}} V^{(8)}_\mu
\end{equation}
with
\begin{eqnarray}
V^{(3)}_\mu & = & \frac{1}{2} \bar{\Psi} \gamma_\mu \lambda^3 \Psi
\nonumber \\
V^{(8)}_\mu & = & \frac{1}{2} \bar{\Psi} \gamma_\mu \lambda^8 \Psi.
\end{eqnarray}
 The electromagnetic form factors $F_i (Q^2)$ can be expressed in terms of
the Sachs form factors, $G_E (Q^2)$ and $G_M (Q^2)$:
\begin{eqnarray}
G^{B}_{E}(Q^2) & = & F^{B}_{1}(Q^2) \;-\;
\frac{Q^2}{4M^{2}_{N}} F^{B}_{2}(Q^2), \nonumber \\
G^{B}_{M} (Q^2) & = & F^{B}_{1} (Q^2)
\;+\; F^{B}_{2} (Q^2) .
\end{eqnarray}
 In the non--relativistic limit($Q^2 << M^{2}_N$),
the Sachs form factors $G_E (Q^2)$ and $G_M (Q^2)$
are related to the time and space components of the electromagnetic
current, respectively:
\begin{eqnarray}
\langle B', p' |V_0(0)| B, p \rangle
& = &  G^{B}_{E} (Q^2) \nonumber \\
\langle B', p' | V_i(0) |B, p \rangle & = &
\frac{1}{2M_N} G^{B}_{M} (Q^2) i
\epsilon_{ijk} q^j \langle \lambda' | \sigma_k | \lambda \rangle,
\label{Eq:gm}
\end{eqnarray}
where $\sigma_k$ stand for Pauli spin matrices.  $| \lambda\rangle$
is the corresponding spin state of the baryon.

The matrix elements of the electromagnetic current can be represented
by the Euclidean functional integral in our model
defined by eq.~(\ref{eq:Z})
\begin{eqnarray}
\langle B', p' | V_\mu(0) | B, p \rangle & = &
\frac{1}{\cal Z} \lim_{T \rightarrow \infty} \exp{(ip_4 \frac{T}{2}
- ip'_{4} \frac{T}{2})} \nonumber \\
& \times & \int d^3 x d^3 y
\exp{(-i \vec{p'} \cdot \vec{y} + i \vec{p} \cdot \vec{x})}
\int {\cal D}U \int {\cal D} \Psi \int {\cal D}\Psi^\dagger
\nonumber \\
& \times & \; J_{B'}(\vec{y},T/2)\Psi^\dagger(0)
\beta \gamma_\mu \hat{Q} \Psi (0) J^{\dagger}_{B} (\vec{x}, -T/2)
\nonumber \\ & \times &
\exp{\left[ - \int d^4 z \Psi^\dagger i D \Psi \right ]}.
\label{Eq:ev}
\end{eqnarray}
 The baryonic states $| B,p\rangle$
and $\langle B', p'|$ are respectively
defined by
\begin{eqnarray}
| B,p \rangle & = & \lim_{x_4 \rightarrow -\infty}
\exp{(ip_4 x_4)}  \frac{1}{\sqrt{Z}}
\int d^3 x \exp{(i\vec{p} \cdot \vec{x})}
J^{\dagger}_{B} (\vec{x},x_4) | 0 \rangle
\nonumber \\
\langle B', p'| & = & \lim_{y_4 \rightarrow +\infty}
\exp{(-ip'_4 y_4)} \frac{1}{\sqrt{Z}}
\int d^3 y \exp{(-i\vec{p'} \cdot \vec{y})}
\langle 0 | J_{B'} (\vec{y}, y_4)
\end{eqnarray}
The baryon current $J_B$ can be constructed from quark fields with the number
of colors $N_c$
\begin{equation}
J_B(x)\;=\; \frac{1}{N_c !} \epsilon_{i_1 \cdots i_{N_c}}
\Gamma^{\alpha_1 \cdots
\alpha_{N_c}}_{JJ_3TT_3Y}\psi_{\alpha_1i_1}(x)
\cdots \psi_{\alpha_{N_c}i_{N_c}}(x).
\end{equation}
$\alpha_1 \cdots\alpha_{N_c}$ denote spin--flavor indices, while
$i_1 \cdots i_{N_c}$ designate color indices.  The matrices
$\Gamma^{\alpha_1 \cdots\alpha_{N_c}}_{JJ_3TT_3Y}$ are taken to endow
the corresponding current with the quantum numbers $JJ_3TT_3Y$.
 The $J^{\dagger}_{B}$ plays the  role of creating the baryon state.
With the quark fields being integrated out,
eq.~(\ref{Eq:ev}) can be divided into two separated contributions:
\begin{equation}
\langle B', p'| V_\mu(0) | B, p \rangle
\;=\;\langle B', p'| V_\mu(0) |B, p \rangle_{val}
\;+\;\langle B', p'| V_\mu(0) |B, p \rangle_{sea},
\end{equation}
where
\begin{eqnarray}
\langle B', p'| V_\mu(0) | B, p \rangle_{val} & = &
\frac{1}{\cal Z}
\Gamma^{\beta_1 \cdots \beta_{N_c}}_{J'J'_3T'T'_3Y'}
\Gamma^{\alpha_1 \cdots \alpha_{N_c}*}_{JJ_3TT_3Y}
\lim_{T \rightarrow \infty} \exp{(ip_4 \frac{T}{2}
- ip'_{4} \frac{T}{2})}
\nonumber \\
& \times & \int d^3x d^3y \exp{(-i\vec{p'}\cdot\vec{y}
+ i \vec{p} \cdot \vec{x})} \nonumber \\
& \times & \int {\cal D}U \exp{(-S_{eff})}
\sum^{N_c}_{i=1} \; _{\beta_i}\langle\vec{y}, {\mbox T}/2|
\frac{1}{i D} | 0,t_z \rangle_{\gamma}
[\beta \gamma_\mu \hat{Q}]_{\gamma \gamma'}
\nonumber \\ & \times &
_{\gamma'}\langle 0,t_z | \frac{1}{i D}
| \vec{x}, -{\mbox T}/2\rangle_{\alpha_i}
\prod^{N_c}_{j \neq i}
\; _{\beta_j} \langle \vec{y},{\mbox T}/2 |
\frac{1}{i D} |\vec{x}, -{\mbox T}/2 \rangle_{\alpha_j}
\label{Eq:val1}
\end{eqnarray}
and
\begin{eqnarray}
\langle B', p'| V_\mu(0) |B, p \rangle_{sea} & = &
\frac{1}{\cal Z}
\Gamma^{\beta_1 \cdots \beta_{N_c}}_{J'J'_3T'T'_3Y'}
\Gamma^{\alpha_1 \cdots \alpha_{N_c}*}_{JJ_3TT_3Y}
\lim_{T \rightarrow \infty} \exp{(ip_4 \frac{T}{2}
- ip'_{4} \frac{T}{2})}
\nonumber \\
& \times & \int d^3x d^3y \exp{(-i\vec{p'}\cdot\vec{y}
+ i \vec{p} \cdot \vec{x})} \nonumber \\
& \times & \int {\cal D}U \exp{(-S_{eff})}
{\rm Tr}\ _{\gamma \lambda c}
\langle 0, t_z | \frac{1}{i D}[\beta \gamma_\mu \hat{Q}]
| 0, t_z \rangle        \nonumber \\
& \times & \prod^{N_c}_{i=1}\; _{\beta_i}
\langle \vec{y},{\mbox T}/2 |   \frac{1}{i D}
| \vec{x}, -{\mbox T}/2\rangle_{\alpha_i}.
\label{Eq:sea1}
\end{eqnarray}
$S_{eff}$ is the effective chiral action expressed by
\begin{equation}
S_{eff} \;=\; -{\mbox{Sp}} \log{\left [ \partial_\tau \;+\; H(U)
\; + \; \beta \hat{m} \;-\;\beta\bar{m}{\bf 1}\right ]}.
\end{equation}
$\mbox{Sp}$ stands for the functional trace of the time--independent
function.

The integral over bosonic fields can be carried out
by the saddle point method in the large $N_c$ limit,
choosing the following Ansatz:
\begin{equation}
U\;=\; \left ( \begin{array}{cc}
U_0 & 0 \\ 0 & 1 \end{array} \right ),
\label{Eq:imbed}
\end{equation}
where $U_0$ is the SU(2) chiral background field
\begin{equation}
U_0\;=\;\exp{[\vec{n}\cdot\vec{\tau}P(r)]} .
\end{equation}
$P(r)$ denotes the profile function satisfying the boundary condition
$P(0) = \pi$ and $P(\infty)=0$.
 In order to find the quantum $1/N_c$ corrections,
we have to integrate eqs.~(\ref{Eq:val1},\ref{Eq:sea1}) over
small oscillations of the pseudo-Goldstone field around
the saddle point eq.~(\ref{Eq:imbed}).  This will not be done
except for the zero modes.  The corresponding fluctuations of the
pion fields are not small and hence cannot be neglected.
 The zero modes are pertinent to continuous symmetries
in our problem.  Actually, there are three translational
and seven rotational zero modes.
 We have to take into account the translational
zero modes properly in order to evaluate form factors,
since the soliton is not invariant under translation and its translational
invariance is restored only after integrating over the translational
zero modes.
 The rotational zero modes determine
the quantum numbers of baryons~\cite{anw}.
 Explicitly, the zero modes are taken into account by
considering a slowly {\em rotating} and {\em translating} hedgehog:
\begin{equation}
\tilde{U}(\vec{x}, t)\;=\; A(t)
U(\vec{x}-\vec{Z}(t)) A^{\dagger} (t).
\label{Eq:rot}
\end{equation}
$A(t)$ belongs to an SU(3) unitary matrix.
 The Dirac operator $i\tilde{D}$ in eq.~(\ref{Eq:Dirac}) can be written
as
\begin{equation}
i \tilde{D} \; = \; \left(\partial_\tau \; + \; H(U)
\;+\;A^{\dagger} (t) \dot{A}(t)
\;-\; i \beta \dot{\vec{Z}} \cdot \nabla
\;+\; \beta A^{\dagger} (t) (\hat{m}-\bar{m}{\bf 1}) A(t) \right).
\end{equation}
The corresponding collective action is expressed by
\begin{eqnarray}
\tilde{S}_{eff} & = & -N_c {\rm Sp}\
\log{\left [ \partial_\tau \;+\; H(U)
\;+\; A^{\dagger} (t) \dot{A}(t)
\;-\;  i \beta \dot{\vec{Z}} \cdot \nabla \right .}
\nonumber \\
&  & \left .
\; + \; \beta A^{\dagger}(t) (\hat{m}-\bar{m}{\bf 1}) A(t)
\;-\; \beta A^{\dagger}(t)
V_\mu \gamma_\mu \hat{Q} A(t) \right ]
\label{Eq:effact}
\end{eqnarray}
with the angular velocity
\begin{equation}
A^{\dagger}(t)\dot{A}(t) \;=\; i\Omega_E \;=\;
\frac{1}{2} i \Omega^{a}_{E} \lambda^a
\end{equation}
and the velocity of the translational motion
\begin{equation}
\dot{\vec{Z}}\;=\; \frac{d}{dt} \vec{Z}
\end{equation}
The canonical quantization of the SU(3) soliton can be found in
Ref.\ \cite{Guad,toy}.
 Expanding eq.~(\ref{Eq:effact}) in powers of
angular and translational velocities ($\sim 1/N_c$),
we end up with the action for collective coordinates:
\begin{equation}
S_{coll} \;\approx \; -N_c {\rm Tr}\ \log{iD} \;+\;
S_{rot}[A]\;+\;S_{trans}[\vec{Z}]
\label{Eq:scoll}
\end{equation}
where
\begin{eqnarray}
S_{rot}[A] & = & \frac{1}{2}  I_{ab} \int dt\Omega_a \Omega_b
\nonumber \\
S_{trans}[\vec{Z}] & = &
\frac{1}{2}  M_{cl} \int dt \dot{\vec{Z}}\cdot \dot{\vec{Z}} ,
\end{eqnarray}
with the moments of inertia $I^{ab}$ calculated in
Ref.\ \cite{betal2}.  $M_{cl}$ is a classical mass of the soliton.
 Corresponding collective hamiltonians have a form:
\begin{eqnarray}
H_{rot} & = &   (I^{-1})_{ab} J_a J_b  , \nonumber \\
H_{trans} & = & \frac{\vec{P}\cdot \vec{P} }{ 2M_{cl}}   ,
\end{eqnarray}
where $J_a$ are
operators of angular momentum and $\vec{P} $  are momentum
operators.

Hence, eq.~(\ref{Eq:val1})
and eq.~(\ref{Eq:sea1}) can be written in terms of
the rotated Dirac operator $i\tilde{D}$ and chiral effective action
$\tilde{S}_{eff}$.
 The functional integral over the pseudoscalar field
$U$ is replaced by the path integral which can be calculated
in terms of the eigenstates of the hamiltonian
corresponding to the collective action given in eq.~(\ref{Eq:scoll})
and these hamiltonians can be diagonalized in an exact manner.
 Therefore, eqs.~(\ref{Eq:val1},\ref{Eq:sea1}) can be rewritten as
ordinary integrals:
\begin{eqnarray}
\langle B', p'| V_\mu(0) | B, p \rangle_{val} & = &
\frac{1}{\cal Z}
\Gamma^{\beta_1 \cdots \beta_{N_c}}_{J'J'_3T'T'_3Y'}
\Gamma^{\alpha_1 \cdots \alpha_{N_c}*}_{JJ_3TT_3Y}
\exp{\left[-((N_c-1) E_{val} + E_{sea})T\right]}
\nonumber \\ & \times &
\lim_{T \rightarrow \infty}
\int d^3x d^3y \exp{(-i\vec{p'}\cdot\vec{y}
\;+ \;i \vec{p} \cdot \vec{x})} \nonumber \\
& \times & \int dA_f dA dA_i d\vec{Z}_f d\vec{Z}d\vec{Z}_i
\langle \vec{Z}_f | \exp{(-H_{trans}T/2)}
|\vec{Z} \rangle
\nonumber \\ & \times &
\langle \vec{Z}| \exp{(-H_{trans}T/2)}
|\vec{Z}_i \rangle
\langle A_f | \exp{(-H_{rot}T/2)} | A \rangle
\nonumber \\ & \times &
\langle A | \exp{(-H_{rot}T/2)} | A_i \rangle
\nonumber \\ & \times &
\sum^{N_c}_{k=1} {\cal T} \left[  _{\beta_k}
\langle \vec{y}-\vec{Z}_f, {\mbox T}/2 | A_f
\frac{1}{i \tilde{D}} |-\vec{Z}\rangle_{\gamma}
[A^\dagger \beta \gamma_\mu \hat{Q}A]
_{\gamma \gamma'}
\right. \nonumber \\ & \times & \left.
_{\gamma'} \langle -\vec{Z}|
\frac{1}{i \tilde{D}}A^{\dagger}_{i}
| \vec{x}-\vec{Z}_i, -{\mbox T}/2 \rangle_{\alpha_k}
\right],
\label{Eq:val2}
\end{eqnarray}
\begin{eqnarray}
\langle B', p' | V_\mu(0) | B, p \rangle_{sea} & = &
\frac{1}{\cal Z}
\Gamma^{\beta_1 \cdots \beta_{N_c}}_{J'J'_3T'T'_3Y'}
\Gamma^{\alpha_1 \cdots \alpha_{N_c}*}_{JJ_3TT_3Y}
\exp{\left[-(N_c E_{val} + E_{sea})T\right]}
\nonumber \\ & \times &
\lim_{T \rightarrow \infty}
\int d^3x d^3y \exp{(-i\vec{p'}\cdot\vec{y}
\;+ \;i \vec{p} \cdot \vec{x})} \nonumber \\
& \times & \int dA_f dA dA_i d\vec{Z}_f d\vec{Z}d\vec{Z}_i
\langle \vec{Z}_f | \exp{(-H_{trans}T/2)}
| \vec{Z} \rangle
\nonumber \\ & \times &
\langle \vec{Z} | \exp{(-H_{trans}T/2)}
| \vec{Z}_i \rangle
\langle  A_f | \exp{(-H_{rot}T/2)}
| A \rangle
\nonumber \\ & \times &
\langle A | \exp{(-H_{rot}T/2)}
| A_i \rangle
\nonumber \\ & \times &
{\cal T} \left[{\rm Tr}\ _{\gamma \lambda c}\;
\langle -\vec{Z} | \frac{1}{i \tilde{D}}
[A^\dagger \beta \gamma_\mu \hat{Q}A]
| -\vec{Z} \rangle
_{\gamma \gamma'} \right].
\label{Eq:sea2}
\end{eqnarray}
${\cal T}[\cdots]$ denotes the time-ordered product of
collective operators.  This is due to the fact that
the functional integral corresponds to the matrix elements
of the time-ordered products of the collective operators.
In particular,
the time-ordering is very significant when we consider the
magnetic form factors (as in case of the axial constants:
see \cite{chetal,bpg2}),
since the spin operator $J^a$ does
not commute with the SU(3) rotational unitary matrix $A(t)$.
As we integrate over zero modes in the final and initial
states, we obtain the translational and rotational corrections
of the classical energies of the soliton from the
effective actions $S_{trans}$ and $S_{rot}$.
Therefore, introducing the spectral representations of the
quark propagator~\cite{dpp} expressed by the eigenfunctions
of the Dirac hamiltonian $H(U)$ and
making use of relations
\begin{equation}
\int d\vec{Z}_i \langle \vec{Z} |
\exp{(-S_{trans})}  |\vec{Z}_i \rangle
f(\vec{x} - \vec{Z}_i)
\smash{\mathop{\longrightarrow}\limits_{T\rightarrow \infty}}
\langle \vec{Z}| \exp{(-S_{trans})}
| \vec{x} \rangle
\int d^3x' f(\vec{x'}),
\end{equation}
\begin{eqnarray}
\Gamma^{\beta_1 \cdots \beta_{N_c}}_{JJ_3TT_3Y}\int d^3\vec{x'}
\prod^{N_c}_{k=1}[A_f \phi (\vec{x'})]_{\beta_k}
& = & \psi^{(8)*}_{(YTT_3)(Y'JJ_3)} (A_f),
\label{Eq:bprod1}  \\
\Gamma^{\alpha_1 \cdots \alpha_{N_c}*}_{JJ_3TT_3 Y}\int d^3\vec{x'}
\prod^{N_c}_{k=1}[\phi^{\dagger} (\vec{x'})A^{\dagger}_i]_{\alpha_k}
& = & \psi^{(8)}_{(YTT_3)(Y'JJ_3)}(A_i) ,
\label{Eq:bprod2}
\end{eqnarray}
\begin{equation}
\langle A | \exp{(-S_{rot})} | A_i \rangle  =
\sum_{{n\atop (YTT_3)}\atop (Y'JJ_3)}
\psi^{(n)}_{(YTT_3)(Y'JJ_3)} (A)
\psi^{(n)*}_{(YTT_3)(Y'JJ_3)} (A_i)
\exp{\left(-\frac{J(J+1)}{2I}{\rm T}\right)},
\end{equation}
we obtain relatively simple expressions:
\begin{eqnarray}
\langle B', p' | V_\mu(0) | B, p \rangle_{val} & = &
N_c \int d^3 Z
\exp{\left(i\vec{q}\cdot \vec{Z}\right)}
\int_{\mbox{\small SU(3)}} dA \psi^{(n)*}_{\mu \nu} (A)
\psi^{(n)}_{\mu' \nu'} (A) \nonumber \\
&\times   & {\cal T} \left [
{\cal F}^{(\Omega^{0})}_{1}(A) \;+\; {\cal F}^{(\Omega^{1})}_{2}(A)
\;+\;{\cal F}^{(m_{s})}_{3}(A)
\right]
\label{Eq:val3}    \\
\langle B', p' | V_\mu(0) | B, p \rangle_{sea} & = &
N_c \int d^3 Z
\exp{\left(i\vec{q}\cdot \vec{Z}\right)}
\int_{\mbox{\small SU(3)}} dA \psi^{(n)*}_{\mu \nu} (A)
\psi^{(n)}_{\mu' \nu'} (A)
\nonumber \\ & \times &
{\cal T} \left[ {\rm Tr}\ 
\langle \vec{Z} | \frac{1}{i\tilde{D}}
\left[A^\dagger \beta \gamma_\mu \hat{Q}A \right]
|\vec{Z} \rangle
\right] .
\label{Eq:sea3}
\end{eqnarray}
Here, we have considered contributions up to the first order
of $\Omega_E$, {\em i.e.} the $1/N_c$ corrections and the linear
corrections of the strange quark mass $m_s$.
 The mixed term $O(m_s/N_c)$ is relatively small, so that it is
neglected~\cite{kbpg}.
 It is performed by the expansion of the propagator $1/i\tilde{D}$
in terms of $\Omega_E$ and $m_s$:
\begin{equation}
\frac{1}{i\tilde D}\;\approx \;
\frac{1}{\partial_\tau + H}
\;+\; \frac{1}{\partial_\tau + H} (-i \Omega_E)
\frac{1}{\partial_\tau + H}
\; + \; \frac{1}{\partial_\tau + H} (-\beta A^{\dagger} \hat{m} A)
\frac{1}{\partial_\tau + H}  .
\label{Eq:exp}
\end{equation}
The collective SU(3) octet wave functions
$\psi^{(n)}_{\mu' \nu'} (A)$ are
identified with the SU(3) Wigner functions
\begin{equation}
\psi^{(n)}_{(YTT_3)(Y'JJ_3)} (A) \;=\;
\sqrt{dim(n)}(-1)^{Y'/2+J_3}
\left [ \langle Y,T,T_3| D^{(n)} (A)
|-Y',J,-J_3 \rangle \right ]^{*}
\label{Eq:wf}
\end{equation}
as eigenstates of the collective rotational hamiltonian.
 The functions ${\cal F}_i(A)$ are defined as
\begin{eqnarray}
{\cal F}^{(\Omega^{0})}_{1}(A) & = &
\langle val | \beta \gamma_\mu \lambda^a
| val \rangle D^{(8)}_{Qa}(A)
\nonumber \\
{\cal F}^{(\Omega^{1})}_{2}(A) & = & -\sum_n\left[
\langle val | \lambda^a
|n \rangle \langle n |
\beta \gamma_\mu \lambda^b | val \rangle
i\Omega^{a}_{E}(A) D^{(8)}_{Qb}(A) \right.
\nonumber \\ & + & \left.
\langle val | \beta \gamma_\mu \lambda^b
|n \rangle \langle n |
\lambda^a |val \rangle D^{(8)}_{Qb}(A)i\Omega^{a}_{E}(A)
\right]\frac{1}{E_{val} - E_n}
\nonumber \\
{\cal F}^{(m_s)}_{3}(A) & = & -(m_0-\bar{m}) \sum_n\left[
\langle val | \beta
|n \rangle \langle n |
\beta \gamma_\mu \lambda^a
| val \rangle D^{(8)}_{Qa}(A)\right.
\nonumber \\ & + & \left.
\langle val | \beta\gamma_\mu \lambda^a
|n \rangle \langle n |
\beta | val \rangle D^{(8)}_{Qa}(A)
\right]\frac{1}{E_{val} - E_n} \nonumber \\  & - &
m_8 \sum_n\left[
\langle val | \beta\lambda^a
|n \rangle \langle n |
\beta \gamma_\mu \lambda^b | val \rangle
D^{(8)}_{8a}(A)D^{(8)}_{Qb}(A)\right.
\nonumber \\ & + & \left.
\langle val | \beta\gamma_\mu \lambda^b
|n \rangle \langle n |
\beta \lambda^a | val \rangle
D^{(8)}_{Qb}(A)D^{(8)}_{8a}(A)
\right]\frac{1}{E_{val} - E_n}
\end{eqnarray}
$D^{(8)}_{Qa}$ is defined as
$\frac{1}{2}  (D^{(8)}_{3a}+\frac{1}{\sqrt{3}} D^{(8)}_{8a})$.
 The collective SU(3) octet wave function in eq.~(\ref{Eq:wf})
satisfies the orthonormality~\cite{swart}
\begin{equation}
\int dA\psi^{(n')*}_{\mu'\nu'} (A)
\psi^{(n)}_{\mu\nu} (A) \;=\;
\delta_{n'n}\delta_{\mu'\mu}\delta_{\nu'\nu}.
\end{equation}
The subscripts $\mu\nu$ of $\psi^{(n)}_{\mu\nu}$
represent $(YTT_3)(Y'JJ_3)$. $(n)$ stands for the irreducible
representation of SU(3).
 $Y'$ is the negative of the right hypercharge
constrained by $Y_R=\frac{N_c B}{3}=1$.
 Since eq.~(\ref{Eq:sea3}), in particular,
its real part diverges, we have to
regularize it.  We employ the well-known proper time regularization
\begin{equation}
{\mbox Re}S_{eff} \;=\; \frac{1}{2} {\mbox Tr}
\int^{\infty}_{0} \frac{du}{u} e^{-uD^{\dagger} D} \phi (u; \Lambda_i)
\label{Eq:propr}
\end{equation}
with
\begin{equation}
\phi (u; \Lambda_i) \;=\;
\sum_i c_i \theta \left(u - \frac{1}{\Lambda^{2}_{i}} \right).
\end{equation}
The cut-off parameter $\phi (u; \Lambda_i)$
is fixed via reproducing the physical
pion decay constant $f_{\pi} = 93\mbox{MeV}$ and other
mesonic properties~\cite{betal2}.
As was done in case of the valence part,
we take into account  the $1/N_c$ and linear $m_s$
corrections(See appendix A for detail).

Making use of the expansion eq.~(\ref{Eq:exp}) and the
SU(3) octet wave functions and employing the
proper-time regularization,
we arrive at
\begin{eqnarray}
\langle B', p' | V_\mu(0) | B, p\rangle_{val} & = &
N_c \langle  D^{(8)}_{Qa} \rangle_B
{\cal P}^{a}_{\mu ; val} (\vec{q})
\nonumber \\
& + &
\frac{N_c}{2} \sum_m
\left\{ \begin{array}{c}
{\rm sign}(E_m)\langle [D^{(8)}_{Qa}, i\Omega^{b}_{E}] \rangle_B
\delta_{\mu i} \\
\langle \{D^{(8)}_{Qa}, i\Omega^{b}_{E}\} \rangle_B
\delta_{\mu 4}\end{array} \right\}
\frac{{\cal Q}^{ab}_{\mu ; val,m} (\vec{q})}
{E_n - E_{val}}
\nonumber \\
& + & \frac{N_c}{2} \sum_m
\langle \{D^{(8)}_{Qa}, i\Omega^{b}_{E}\} \rangle_B
\delta_{\mu i}
\frac{{\cal Q}^{ab}_{\mu ;val,m} (\vec{q})}
{E_n - E_{val}}
\nonumber \\
& + & N_c (m_0-\bar{m}) \sum_m \langle
D^{(8)}_{Qa} \rangle_B \delta_{\mu i}
\frac{{\cal M}^{a}_{\mu ; val,m} (\vec{q})}
{E_n - E_{val}}
\nonumber \\
& + &
N_c m_8 \sum_m
\langle \{D^{(8)}_{Qa}, D^{(8)}_{8b}\}\rangle_B
\frac{{\cal K}^{ab}_{\mu ; val,m} (\vec{q})}
{E_n - E_{val}}
\nonumber \\
\label{Eq:emf1}
\end{eqnarray}
\begin{eqnarray}
\langle B', p' | V_\mu(0) | B, p \rangle_{sea} & = &
-\frac{N_c}{2} \sum_m {\rm sign} (E_n)
\langle D^{(8)}_{Qa}\rangle_B
\left\{\begin{array}{c} {\cal R} (E_n) \delta_{\mu i} \\
\delta_{\mu 4} \end{array} \right\}
{\cal P}^{a}_{\mu ; n} (\vec{q})
\nonumber \\ & + &
\frac{N_c}{4} \sum_{n,m}
\left\{ \begin{array}{c}
{\cal R}_{\cal Q} (E_n,E_m)
\langle [D^{(8)}_{Qa}, i\Omega^{b}_{E}]\rangle_B
\delta_{\mu i} \\
{\cal R}_{I}(E_n,E_m)
\langle \{D^{(8)}_{Qa}, i\Omega^{b}_{E}\} \rangle_B
\delta_{\mu 4}\end{array} \right\}
{\cal Q}^{ab}_{\mu ; nm} (\vec{q})
 \nonumber \\
& + & \frac{N_c}{4} \sum_{n,m}
\langle \{D^{(8)}_{Qa}, i\Omega^{b}_{E}\} \rangle_B
{\cal R}_{\cal M} (E_n, E_m) \delta_{\mu i}
{\cal Q}^{ab}_{\mu ; nm} (\vec{q})
\nonumber \\
& + & \frac{N_c}{2} (m_0-\bar{m}) \sum_{n,m} \langle
D^{(8)}_{Qa}\rangle_B {\cal R}_{\beta} (E_n, E_m)
{\cal M}^{a}_{\mu ; nm} (\vec{q}) \delta_{\mu i}
\nonumber \\
& + & \frac{N_c}{2} m_8 \sum_{n,m}
\langle \{D^{(8)}_{Qa}, D^{(8)}_{8b}\} \rangle_B
\left\{\begin{array}{c}
{\cal R}_{\beta} (E_n, E_m) \delta_{\mu i}  \\
{\cal R}_{\cal M} (E_n, E_m) \delta_{\mu 4}
\end{array} \right \}
{\cal K}^{ab}_{\mu ; nm} (\vec{q}) ,
\label{Eq:emf2}
\end{eqnarray}
where the quark matrix elements are written as
\begin{eqnarray}
{\cal P}^{a}_{\mu ; n} (\vec{q}) & = &
\int d^3 x e^{i\vec{q} \cdot \vec{x}}
\Psi^{\dagger}_{n} (x)\beta \gamma_\mu
\lambda^a \Psi_{n} (x),\nonumber \\
{\cal Q}^{ab}_{\mu ; nm} (\vec{q}) & = &
\int d^3 x e^{i\vec{q} \cdot \vec{x}}
\int d^3y \Psi^{\dagger}_{n}(x) \beta \gamma_\mu \lambda^a
\Psi_{m}(x)\Psi^{\dagger}_{m}(y)
\lambda^b \Psi_{n}(y),
\nonumber \\
{\cal M}^{a}_{\mu ; nm}(\vec{q}) & = &
\int d^3 x e^{i\vec{q} \cdot \vec{x}}
\int d^3y \Psi^{\dagger}_{n}(x) \beta \gamma_\mu \lambda^a
\Psi_{m}(x)\Psi^{\dagger}_{m}(y)
\beta \Psi_{n}(y),
\nonumber \\
{\cal K}^{ab}_{\mu ; nm}(\vec{q}) & = &
\int d^3 x e^{i\vec{q} \cdot \vec{x}}
\int d^3y\Psi^{\dagger}_{n}(x)
\beta \gamma_\mu \lambda^a
\Psi_{m}(x)\Psi^{\dagger}_{m}(y)
\beta \lambda^b \Psi_{n}(y).
\end{eqnarray}
The regularization functions are given by
\begin{eqnarray}
{\cal R} (E_n) & = & \int \frac{du}{\sqrt{\pi u}}
\phi (u;\Lambda_i) |E_n| e^{-uE^{2}_{n}},
\nonumber \\
{\cal R}_{\cal Q} (E_n, E_m) & = & \frac{1}{2\pi} c_i
\int^{1}_{0} d\alpha \frac{\alpha (E_n + E_m) - E_m}
{\sqrt{\alpha ( 1 - \alpha)}}
\frac{\exp{\left (-[\alpha E^{2}_n + (1-\alpha)E^{2}_m]/
\Lambda^{2}_i  \right)}}{\alpha E^{2}_n + (1-\alpha)E^{2}_m},
\nonumber \\
{\cal R}_{I} (E_n, E_m) & = & - \frac{1}{2\sqrt{\pi}}
\int^{\infty}_{0} \frac{du}{\sqrt{u}} \phi (u;\Lambda_i)
\left [ \frac{E_n e^{-u E^{2}_{n}} +  E_m e^{-u E^{2}_{m}}}
{E_n + E_m} \;+\; \frac{e^{-u E^{2}_{n}} - e^{-u E^{2}_{m}}}
{u(E^{2}_{n} - E^{2}_{m})} \right ],
\nonumber \\
{\cal R}_{\cal M} (E_n, E_m) & = &
\frac{1}{2}  \frac{ {\rm sign} (E_n)
- {\rm sign} (E_m)}{E_n - E_m},
\nonumber \\
{\cal R}_{\beta}  (E_n, E_m) & = &
\frac{1}{2\sqrt{\pi}} \int^{\infty}_{0}
\frac{du}{\sqrt{u}} \phi (u;\Lambda_i)
\left[ \frac{E_n e^{-uE^{2}_{n}} - E_m e^{-uE^{2}_{m}}}
{E_n - E_m}\right].
\label{Eq:regul}
\end{eqnarray}
$I_i$ are moments of inertia defined in Ref.\ \cite{betal1}.
$\langle \rangle_B$ denotes the expectation value of the
Wigner $D$ functions in collective space spanned by $A$.
 The expectation values of the $D$ functions can be evaluated
by SU(3) Clebsch--Gordan coefficients listed
in \cite{swart,mf}. The index $\mu$ is the Lorentz index and
$a$ and $b$ denote the flavors, whereas $i$ designates the
space component of the electromagnetic current.
We can here notice that in eq.~(\ref{Eq:emf2}) $1/N_c$ term includes
two different commuting relations {\em i.e.},
the commutator and anti-commutator between the
SU(3) Wigner function $D^{(8)}$ and
the angular velocity $\Omega_E$ of the soliton.
This is due to the time-ordering of the operators and the
symmetric properties of the quark matrix elements under indices
$n$ and $m$ or under $G^{\gamma_5}$-parity~\cite{pobyl}.
If the quark matrix elements are antisymmetric, then
the commutator survives, while if they are symmetric,
then the anti-commutator does.
The quark matrix elements for the electric form factors ($\mu = 4$)
are symmetric whereas some of
the matrix elements for the magnetic form factors
are anti-symmetric.  However, note that on the whole the matrix element
of the current is symmetric, since the regularization functions
are symmetric under exchange of $n$ and $m$ except for
${\cal R}_{\cal Q}$.

The regularization functions in eq.~(\ref{Eq:regul})
are determined in the proper time regularization manifestly
except for ${\cal R}_{\cal M}$ which corresponds to
the Wess-Zumino terms from the imaginary part of the action.
In fact, ${\cal  R}_{\cal M}$ is not a regularization function.
It is independent of the cut-off parameter $\Lambda$.

With SU(3) symmetry explicitly broken by $m_s$, the collective
hamiltonian is no longer SU(3)-symmetric.  Therefore,
the eigenstates of the hamiltonian are not in a pure
octet or decuplet but mixed states.
Treating $m_s$ as a perturbation,
we can obtain the mixed SU(3) baryonic states:
\begin{equation}
| 8, B \rangle \;=\; | 8,B \rangle \;+ \;
c^{B}_{\bar{10}} | \bar{10},B \rangle
\;+\;c^{B}_{27} | 27,B \rangle
\label{Eq:wfc}
\end{equation}
with
\begin{equation}
c^{B}_{\bar{10}} \;=\; \frac{\sqrt{5}}{15}(\sigma - r_1)
\left[ \begin{array}{c}  1 \\ 0 \\ 1 \\ 0
\end{array} \right] I_2 m_s,
c^{B}_{27} \;=\; \frac{1}{75}(3\sigma + r_1 - 4r_2)
\left[ \begin{array}{c} \sqrt{6} \\ 3 \\ 2 \\  \sqrt{6}
\end{array}\right] I_2 m_s.
\label{Eq:g2}
\end{equation}
in the basis [$N$, $\Lambda$, $\Sigma$, $\Xi$].
Here, $B$ denotes the SU(3) octet baryons with the spin 1/2.
The constant $\sigma$ is related to the SU(2) $\pi N$ sigma term
$\Sigma_{SU(2)}\;=\;3/2 (m_u + m_d) \sigma$ and $r_i$ designates
$K_i/I_i$, where $K_i$ stands for the anomalous moments of inertia
defined in~Ref.\ \cite{betal2}.

\section{The electric properties of the SU(3) octet baryons}
The electric form factors are easily obtained by the
matrix elements of the time component of the electromagnetic
current, as was defined in eq.~(\ref{Eq:gm}).
eq.~(\ref{Eq:emf2}) furnishes the final expression of the
electric form factor.  Since the SU(3) hedgehog solutions
are obtained by means of the embedding of the SU(2)
hedgehog field $U_0$ as shown in~eq.~(\ref{Eq:imbed}),
it is convenient to
define the projection operators $P_{T}$ and $P_S$
\begin{equation}
P_T\;=\; \left( \begin{array}{ccc}
1 & 0 & 0 \\  0 & 1 & 0 \\ 0 & 0 & 0 \end{array} \right)
,\;\; P_S  \;=\; \left( \begin{array}{ccc}
0 & 0 & 0 \\  0 & 0 & 0 \\ 0 & 0 & 1 \end{array} \right).
\label{Eq:proj}
\end{equation}
Having defined these projection operators, we can separate
the pure SU(2) part from the SU(3) which are represented by
the collective operators.  Utilizing the projection
operators and introducing
${\mbox{SU(2)}}_{\rm T} \times {\mbox{U(1)}}_{\rm Y}$
invariant tensors
\begin{eqnarray}
P_T \lambda^a & = & \left\{ \begin{array}{ll}
\tau^a & \mbox{if $a=1,2,3$} \\
0      & \mbox{if $a=4,5,6,7$}\\
1      & \mbox{if $a=8$}
\end{array} \right.
\nonumber   \\
P_T \lambda^a P_S \lambda^b
& = & \left[i(f^{abc} - \epsilon^{abc})
- \frac{1}{\sqrt{3}}
(\delta^{ac} \delta^{b8} + \delta^{a8} \delta^{bc})
+ d^{abc}\right]\lambda^c,
\label{Eq:itensor}
\end{eqnarray}
we can find that the quark matrix elements include only
the pure SU(2) components with transition matrix elements
between the vacuum states with SU(2) flavors
and the eigenstates of the
one-body Hamiltonian~eq.~(\ref{Eq:hamil}).
The SU(3) elements only appear in the collective parts.
Hence, we can write the expression of the electric form factors
\begin{eqnarray}
G^{B}_{E} (\vec{Q}^2) & = &
\frac{N_c}{\sqrt{3}}
\langle D^{(8)}_{Q8} \rangle_B {\cal B}(\vec{Q}^2) \;-\;
\langle D^{(8)}_{Qa}J_a \rangle_B
\frac{2{\cal I}_1(\vec{Q}^2)}{I_1}\;-\;
\langle D^{(8)}_{Qp}J_p \rangle_B
\frac{2{\cal I}_2(\vec{Q}^2)}{I_2} \nonumber \\
& + &
\langle D^{(8)}_{8a} D^{(8)}_{Qa} \rangle_B
\frac{4 m_s}{I_1\sqrt{3}} \left (I_1 {\cal K}_1 (\vec{Q}^2)
-{\cal I}_1 (\vec{Q}^2) K_1\right ) \nonumber \\
& + &
\langle D^{(8)}_{8p} D^{(8)}_{Qp} \rangle_B
\frac{4 m_s}{I_2\sqrt{3}} \left (I_2 {\cal K}_2 (\vec{Q}^2)
-{\cal I}_2 (\vec{Q}^2) K_2 \right ),
\label{Eq:elecf}
\end{eqnarray}
where
\begin{eqnarray}
{\cal B}(\vec{Q}^2) & = & \int d^3 x \; j_0 (Qr)
\left [ \Psi^{\dagger}_{val}(x) \Psi_{val} (x) \;-\;\frac{1}{2} \sum_n
{\rm sign} (E_n) \Psi^{\dagger}_{n}(x)
\Psi_{n} (x) \right ], \nonumber \\
{\cal I}_1 (\vec{Q}^2) & = & \frac{N_c}{6} \sum_{n, m}
\int d^3 x \;j_0 (Qr) \int d^3 y
\left [\frac{\Psi^{\dagger}_{n} (x) \vec{\tau} \Psi_{val} (x) \cdot
\Psi^{\dagger}_{val} (y) \vec{\tau} \Psi_{n} (y)}
{E_n - E_{val}} \right .
\nonumber \\  & & \hspace{3cm} \;+\; \left . \frac{1}{2}
\Psi^{\dagger}_{n} (x) \vec{\tau} \Psi_{m} (x) \cdot
\Psi^{\dagger}_{m} (y) \vec{\tau} \Psi_{n} (y)
{\cal R}_{\cal I} (E_n, E_m) \right ],
\nonumber \\
{\cal I}_2 (\vec{Q}^2) & = &\frac{N_c}{6} \sum_{n, m^{0}}
\int d^3 x \;j_0 (Qr)\int d^3 y
\left [\frac{\Psi^{\dagger}_{m^{0}} (x) \Psi_{val} (x)
\Psi^{\dagger}_{val} (y) \Psi_{m{^0}} (y)}
{E_{m^{0}} - E_{val}} \right .
\nonumber \\  & & \hspace{3cm} \;+\;\left . \frac{1}{2}
\Psi^{\dagger}_{n} (x) \Psi_{m^{0}} (x)
\Psi^{\dagger}_{m^{0}} (y) \Psi_{n} (y)
{\cal R}_{\cal I} (E_n, E_m^{0}) \right ],
\nonumber \\
{\cal K}_1 (\vec{Q}^2) & = & \frac{N_c}{6} \sum_{n, m}
\int d^3 x \;j_0 (Qr) \int d^3 y
\left [\frac{\Psi^{\dagger}_{n} (x) \vec{\tau} \Psi_{val} (x) \cdot
\Psi^{\dagger}_{val} (y) \beta \vec{\tau} \Psi_{n} (y)}
{E_n - E_{val}} \right .
\nonumber \\  & & \hspace{3cm} \;+\; \left . \frac{1}{2}
\Psi^{\dagger}_{n} (x) \vec{\tau} \Psi_{m} (x) \cdot
\Psi^{\dagger}_{m} (y) \beta \vec{\tau} \Psi_{n} (y)
{\cal R}_{\cal M} (E_n, E_m)
\right ],
\nonumber \\
{\cal K}_2 (\vec{Q}^2) & = & \frac{N_c}{6} \sum_{n, m^{0}}
\int d^3 x \;j_0 (Qr) \int d^3 y
\left [\frac{\Psi^{\dagger}_{m^{0}} (x) \Psi_{val} (x)
\Psi^{\dagger}_{val} (y) \beta \Psi_{m{^0}} (y)}
{E_{m^{0}} - E_{val}} \right .
\nonumber \\  & & \hspace{3cm} \;+\;\left . \frac{1}{2}
\Psi^{\dagger}_{n} (x) \Psi_{m^{0}} (x)
\Psi^{\dagger}_{m^{0}} (y) \beta \Psi_{n} (y)
{\cal R}_{\cal M} (E_n, E_m^{0}) \right ]
\end{eqnarray}
with the regularization functions
${\cal R}_{I}$ and ${\cal R}_{\cal M}$ defined in
Ee.~(\ref{Eq:regul}).
The subscripts $a$ and $p$ denote
the flavor indices $a=1,2,3$ and $p=4,\cdots 7$, respectively,
and $m^0$ denotes the vacuum state with the SU(2) flavor.
$j_0 (Qr)$ is the spherical Bessel function of integral order 0.
We can see that when $\vec{Q}^2=0$ ${\cal B}$ becomes
the baryon number $B=1$, while ${\cal I}_i$ and ${\cal K}_i$ become
the usual and the anomalous moments of inertia,
respectively.  In that case,
eq.~(\ref{Eq:elecf}) is reduced to the Gell-Mann--Nishjima formula
$Q = T_3 + \frac{1}{2}  Y$,
using the relation
\begin{equation}
\sum^8_{a=1} D^{(8)}_{3a} R^a \;=\; L_3 = T_3,\;\;\;
\sum^8_{a=1} D^{(8)}_{8a} R^a \;=\; L_8 = \frac{1}{2}  \sqrt{3} Y .
\end{equation}
At $\vec{Q}^2 = 0$, the mass corrections do not contribute to
the electric form factors, since the fourth and fifth terms in
eq.~(\ref{Eq:elecf}) vanish at the zero momentum transfer.

In order to calculate the form factors and other
observables numerically, we follow the
well--known Kahana and Ripka method~\cite{rk}.
Since the isovector electric charge radii have
a pole in the chiral limit, we take the
pion mass $m_\pi=139\; \rm{MeV}$ into account.
The self-consistent profile function obtained by the
Kahana-Ripka method has a good behavior in the solitonic
region, but the tail of the pion field is spoiled a little
due to the finite size of the radial box
when we take into account the pion mass.
Hence, at large distances we use the exact Yukawa-type
asymptotic behavior of the profile function:
\begin{equation}
P(r) \;=\; \alpha \exp{(-m_\pi r)} \frac{1 + m_\pi r}{r^2},
\label{Eq:pion}
\end{equation}
where $\alpha$ is a constant governing the strength of the
pion field.  It is determined by matching the self-consistent
profile function to the aymptotic tail
given in eq.~(\ref{Eq:pion}) at large distances, {\em i.e.} about 4 fm.
Since the neutron electric form factor,
electromagnetic charge radius and magnetic form factors are
very sensitive to the long-range tail, we have to use the larger
size of the radial box.  Hence, we employ the box size
$D\simeq 10 \;\mbox{fm}$ which is large enough to incorporate
the long-range part properly.

Figure 1 shows the electric form factor of the proton while
Figure 2 draws that of the neutron as a function
of $Q^2$ with the constituent quark mass 370 MeV,
420 MeV and 450 MeV.  The empirical data are provided by
Ref.~\cite{pl}.
 From Figure 1, we can easily find that
the proton electric form factor $(G^{p}_{E})$ increases as the constituent
quark mass does.  For the best fit, we choose
the constituent quark mass $M=420\mbox{MeV}$ as usually done
for the other observables.  However, the neutron electric form
factor $(G^{n}_{E})$ does not show such dependence
on the $M$ as that of the proton
does.  The dependence of $G^{n}_{E}$ is not monotonous.
 As shown in Figure 2,
$G^{n}_{E}$ with $M=420\;\mbox{MeV}$
is greater than those in the case of $M=370\;\mbox{MeV}$ and
$M=450\;\mbox{MeV}$.  At the first glance, it might seem to
be strange.  However, since $G^{n}_{E}$ is
a very tiny and sensitive quantity, one should carefully examine
each contribution to it.  Having scrutinized each contribution,
we find that the wave function corrections
given by eq.(\ref{Eq:wfc}) are responsible for the above-mentioned
behavior in $G^{n}_{E}$.  In particular,
the $\sigma$ appearing in eq.(\ref{Eq:g2})
plays a pivotal role of governing the
behavior of $G^{n}_{E}$.
As $M$ increases, the electric form factors increase but
the $\sigma$ decreases.  In the meanwhile, the $G^{n}_{E}$
gets an optimal value around $420\;\mbox{MeV}$.

The contribution of the $m_s$ corrections with the wave
function corrections is displayed in Figure 3-4.
 In fact, the $m_s$ corrections
without the collective wave functions modified bring
$G^{n}_{E}$ down sizably, since the $m_s$ terms
$(I_i {\cal K}_i (\vec{Q}^2) -{\cal I}_i (\vec{Q}^2) K_i)$
diminish electric form factors in general.  However,
as explained above, the collective wave function corrections
are in particular significant in order to improve $G^{n}_{E}$.
 On the contrary to the case of the $G^{p}_{E}$ to which the
wave function corrections contribute about $1\%$, those
contributions to $G^{n}_{E}$ are strong enough to overcome
the $m_s$ corrections.  As a result, the total $m_s$ corrections
enhance $G^{n}_{E}$ about $20\%\sim 30\%$ in the small $Q^2$ region.

More important observables for us are probably electric charge radii
which are determined by the behavior of the electric form factors
near $Q^2=0$ which are defined by
\begin{equation}
\langle r^2 \rangle^{B}_{E} \;=\; -6 \left.
\frac{dG^{B}_{E} (Q^2)}{dQ^2} \right|_{Q^2=0}.
\label{Eq:elecr}
\end{equation}
Using eq.~(\ref{Eq:elecr}), we obtain the electric charge radii of
the proton and the neutron
$\langle r^2 \rangle^{th}_{p} = 0.78\;{\mbox fm}^2$ and
$\langle r^2 \rangle^{th}_{n} = -0.09 \; {\mbox fm}^2$, respectively.
The experimental data are
$\langle r^2\rangle_p = 0.74 \;{\mbox fm}^2 $ and
$\langle r^2 \rangle_n = -0.113\pm 0.003\; {\mbox fm}^2 $
\cite{Kopeckyetal}.  We can see that our results
are in a good agreement with experimental ones within
about 10$\%$.

In dotted curves in Figures 3-4 we show the prediction of the SU(2)
model~\cite{chetal}.
As for the proton electric form factor,
it is comparable to the SU(3), whereas
a great discrepancy is observed in case of the neutron electric
form factor.  This discrepancy can be understood by
looking into the electric isospin form factors.  Figure 5 shows
differences in the electric isospin form factors
between the SU(2) and SU(3) model.  From Figure 5,
we can find that in case of the SU(3)
the difference between the isoscalar
and isovector form factors are quite small while their sum is
comparable.  The discrepancy in the neutron form factors
lies in this difference between electric isospin form factors.
It is partly due to the absence of $m_s$ and terms depending
on the $I_2$ in the SU(2) model
and partly due to the different expectation values of the
collective operators.
 In particular, the terms with
the $I_2$ in eq.~(\ref{Eq:elecf}) can be understood as kaonic
contributions in the mesonic language~\cite{wkg}.
They are relevant to the hidden strangeness having an effect
on the nucleon.

We now turn our attention to the other SU(3) hyperons.
In Figures 6-7 we present the electric form factors for the
SU(3) octet hyperons.  Figure 6 draws those of charged hyperons
while Figure 7 displays those of neutral ones.
Without $m_s$ correction, we could observe
$U$-spin symmetry expressed by
\begin{eqnarray}
G^{p}_{E,M} & = & G^{\Sigma^{+}}_{E,M},\;\;
G^{\Sigma^{-}}_{E,M}=G^{\Xi{-}}_{E,M}, \nonumber \\
G^{n}_{E,M} & = & G^{\Xi^{0}}_{E,M},\;\;
G^{\Lambda}_{E,M} \;=\; -G^{\Sigma^{0}}_{E,M}.
\label{Eq:uspin}
\end{eqnarray}
Figures 6-7 show us SU(3) symmetry
breaking arising from the $m_s$ correction.
In case of the charged octet baryons the SU(3) splitting of
the electric form factors are rather small while
it is quite visible for the neutral ones.
The predicted electric charge radii for different baryons
are listed in table 1, compared with the SU(3)
Skyrme model with pseudoscalar vector mesons~\cite{pw}.
\section{The magnetic properties of the SU(3) octet baryons}
The space components of the electromagnetic current is
responsible for the magnetic form factors.
As used in case of the electric form factor,
we again make use of the projection operators
given in eq.~(\ref{Eq:proj})
and ${\mbox{SU(2)}}_{\rm T} \times {\mbox{U(1)}}_{\rm Y}$
invariant tensors, so that we
obtain the expression of $G^{B}_{M} (\vec{Q}^2)$:
\begin{eqnarray}
G^{B}_{M} (\vec{Q}^2) & = & \frac{M_N}{|\vec{Q}|}
\left [\langle D^{(8)}_{Q3}\rangle_B \left({\cal Q}_0 (\vec{Q}^2)
\; +\; \frac{{\cal Q}_1(\vec{Q}^2)}{I_1}
\; +\; \frac{{\cal Q}_2(\vec{Q}^2)}{I_2}\right) \right.
\nonumber \\
&-& \langle D^{(8)}_{Q8}J_3 \rangle_B
\frac{{\cal X}_1 (\vec{Q}^2)}{\sqrt{3}I_1}
\;-\;   \langle d_{3pq}D^{(8)}_{Qp}J_q \rangle_B \delta_{pq}
\frac{{\cal X}_2 (\vec{Q}^2)}{I_2}
\nonumber \\
&+& 2m_s \langle (D^{(8)}_{88} - 1)D^{(8)}_{Q3} \rangle_B
{\cal M}_0 (\vec{Q}^2) \nonumber \\
& + &  m_s  \langle D^{(8)}_{83}D^{(8)}_{Q8} \rangle_B
\left(2 {\cal M}_1 (\vec{Q}^2)  \;-\;
\frac{2}{3} r_1 {\cal X}_1 (\vec{Q}^2)  \right)
\nonumber \\
& + & \left . m_s  \sqrt{3}
\langle d_{3pq}D^{(8)}_{8p}D^{(8)}_{Qq} \rangle_B\delta_{pq}
\left(2 {\cal M}_2 (\vec{Q}^2)
\; - \;
\frac{2}{3} r_2 {\cal X}_2 (\vec{Q}^2)  \right) \right ],
\label{Eq:magf}
\end{eqnarray}
where
\begin{eqnarray}
{\cal Q}_0 (\vec{Q}^2) & = &
{N_c}\int d^3 x j_1 (qr)
\left[ \Psi ^{\dagger}_{val}(x) \gamma_{5}
\{\hat{r} \times \vec{\sigma} \} \cdot \vec{\tau}
\Psi_{val} (x) \right. \nonumber \\ & &
\left . \hspace{1cm} \;-\;
\frac{1}{2}  \sum_n {\rm sign} (E_n)
\Psi ^{\dagger}_{n}(x) \gamma_{5}
\{\hat{r} \times \vec{\sigma} \} \cdot \vec{\tau}
\Psi_{n}(x) {\cal R}(E_n)\right ],
\nonumber \\
{\cal Q}_1 (\vec{Q}^2) & = &  \frac{iN_c}{2}\sum_{n}
\int d^3 x j_1 (qr)
\int d^3 y \nonumber \\  & \times &
\left[{\rm sign} (E_n)
\frac{\Psi^{\dagger}_{n} (x) \gamma_{5}
\{\hat{r} \times \vec{\sigma} \} \times \vec{\tau}
\Psi_{val} (x) \cdot
\Psi^{\dagger}_{val} (y) \vec{\tau} \Psi_{n} (y)}
{E_n - E_{val}} \right .
\nonumber \\  & & 
\;+\; \left . \frac{1}{2} \sum_{m}
\Psi^{\dagger}_{n} (x)\gamma_{5} \{\hat{r} \times \vec{\sigma} \}
\times \vec{\tau}  \Psi_{m} (x) \cdot
\Psi^{\dagger}_{m} (y) \vec{\tau} \Psi_{n} (y)
{\cal R}_{\cal Q} (E_n, E_m) \right ],
\nonumber \\
{\cal Q}_2 (\vec{Q}^2) & = &  \frac{N_c}{2} \sum_{m^0}
\int d^3 x j_1 (qr)
\int d^3 y \nonumber \\  & \times &
\left[ {\rm sign} (E_{m^0})
\frac{\Psi^{\dagger}_{m^0} (x) \gamma_{5}
\{\hat{r} \times \vec{\sigma} \} \cdot \vec{\tau}
\Psi_{val} (x)
\Psi^{\dagger}_{val} (y) \Psi_{m^0} (y)}
{E_{m^0} - E_{val}} \right .
\nonumber \\  & + & 
\left . \sum_{n}
\Psi^{\dagger}_{n} (x)\gamma_{5} \{\hat{r} \times \vec{\sigma} \}
\cdot \vec{\tau}  \Psi_{m^0} (x)
\Psi^{\dagger}_{m^0} (y) \Psi_{n} (y)
{\cal R}_{\cal Q} (E_n, E_{m^0}) \right ],
\nonumber \\
{\cal X}_1 (\vec{Q}^2) & = & N_c
\sum_{n} \int d^3 x j_1 (qr)
\int d^3 y \left[
\frac{\Psi^{\dagger}_{n} (x)\gamma_{5}
\{\hat{r} \times \vec{\sigma} \}
\Psi_{val} (x) \cdot
\Psi^{\dagger}_{val} (y) \vec{\tau} \Psi_{n} (y)}
{E_n - E_{val}} \right .
\nonumber \\  &+ & \left. 
\frac{1}{2} \sum_{m}
\Psi^{\dagger}_{n} (x)\gamma_{5} \{\hat{r} \times \vec{\sigma} \}
\Psi_{m} (x) \cdot
\Psi^{\dagger}_{m} (y) \vec{\tau} \Psi_{n} (y)
{\cal R}_{\cal M} (E_n, E_m) \right ],
\nonumber \\
{\cal X}_2 (\vec{Q}^2) & = & N_c
\sum_{m^0} \int d^3 x j_1 (qr)
\int d^3 y \left[
\frac{\Psi^{\dagger}_{m^0} (x)\gamma_{5}
\{\hat{r} \times \vec{\sigma} \} \cdot \vec{\tau}
\Psi_{val} (x)
\Psi^{\dagger}_{val} (y) \Psi_{m^0} (y)}
{E_{m^0} - E_{val}} \right .
\nonumber \\  & & \hspace{1cm}
\;+\; \left . \sum_{n}
\Psi^{\dagger}_{n} (x)\gamma_{5} \{\hat{r} \times \vec{\sigma} \}
\cdot \vec{\tau} \Psi_{m^0} (x)
\Psi^{\dagger}_{m^0} (y) \Psi_{n} (y)
{\cal R}_{\cal M} (E_n, E_{m^0}) \right ],
\nonumber \\
{\cal M}_0 (\vec{Q}^2) & = &  \frac{N_c}{3} \sum_{n}
\int d^3 x j_1 (qr)
\int d^3 y \left[
\frac{\Psi^{\dagger}_{n} (x) \gamma_{5}
\{\hat{r} \times \vec{\sigma} \} \cdot \vec{\tau}
\Psi_{val} (x)
\Psi^{\dagger}_{val} (y) \beta \Psi_{n} (y)}
{E_{n} - E_{val}} \right .
\nonumber \\  & & \hspace{1cm}
\;+\; \left . \frac{1}{2} \sum_{m}
\Psi^{\dagger}_{n} (x)\gamma_{5} \{\hat{r} \times \vec{\sigma} \}
\cdot \vec{\tau}  \Psi_{m} (x)
\Psi^{\dagger}_{m} (y)\beta \Psi_{n} (y)
{\cal R}_{\beta} (E_n, E_m) \right ],
\nonumber \\
{\cal M}_1 (\vec{Q}^2) & = & \frac{N_c}{3}
\sum_{n} \int d^3 x j_1 (qr)
\int d^3 y \nonumber \\  & \times &
\left[
\frac{\Psi^{\dagger}_{n} (x)\gamma_{5}
\{\hat{r} \times \vec{\sigma} \}
\Psi_{val} (x) \cdot
\Psi^{\dagger}_{val} (y) \beta \vec{\tau} \Psi_{n} (y)}
{E_n - E_{val}} \right .
\nonumber \\  &+ & 
\left . \frac{1}{2} \sum_{m}
\Psi^{\dagger}_{n} (x)\gamma_{5} \{\hat{r} \times \vec{\sigma} \}
\Psi_{m} (x) \cdot
\Psi^{\dagger}_{m} (y) \beta \vec{\tau} \Psi_{n} (y)
{\cal R}_{\beta} (E_n, E_m) \right ],
\nonumber \\
{\cal M}_2 (\vec{Q}^2) & = &  \frac{N_c}{3} \sum_{m^0}
\int d^3 x j_1 (qr)
\int d^3 y \nonumber \\  & \times &
\left[
\frac{\Psi^{\dagger}_{m^0} (x) \gamma_{5}
\{\hat{r} \times \vec{\sigma} \} \cdot \vec{\tau}
\Psi_{val} (x)
\Psi^{\dagger}_{val} (y)\beta \Psi_{m^0} (y)}
{E_{m^0} - E_{val}} \right .
\nonumber \\  &+ & 
\left . \sum_{n}
\Psi^{\dagger}_{n} (x)\gamma_{5} \{\hat{r} \times \vec{\sigma} \}
\cdot \vec{\tau}  \Psi_{m^0} (x)
\Psi^{\dagger}_{m^0} (y) \beta \Psi_{n} (y)
{\cal R}_{\beta} (E_n, E_{m^0}) \right ]  .
\label{Eq:mdens}
\end{eqnarray}
The regularization functions
${\cal R}$, ${\cal R}_{\cal Q}$, ${\cal R}_{\cal M}$
and ${\cal R}_{\beta}$ are defined in eq.~(\ref{Eq:regul}).
The subscripts $p$ and $q$ in eq.~(\ref{Eq:magf})
designate flavor indices from 4 to 7.
The $m^0$ in the summation stands for the vacuum states
with the SU(2) flavor.
$r_i$ is $K_i/I_i$ for short.
As we can see from the densities for the magnetic form factors
in eq.~(\ref{Eq:mdens}), they are pure SU(2) quantities.
The SU(3) components are only found in the collective
operators in eq.~(\ref{Eq:magf}).  Therefore, it is
straightforward to calculate eq.~(\ref{Eq:magf}) numerically.
To make sure, we have compared
the density of each contribution with the corresponding density
in the gradient expansion given in appendix B.
As the soliton size increases, our expressions converge
to those of the gradient expansion.

The nucleon magnetic form factors are displayed in Figures 8-9,
as the constituent quark mass is varied from
$M=370$ to $M=450\mbox{MeV}$.
In contrast to the case of the electric form factors,
the dependence of the magnetic form factors
on the constituent quark mass is not linear.
Up to around $Q^2=0.2\mbox{GeV}^2$ in case
of the proton ($Q^2=0.4\mbox{GeV}^2$ for the
neutron), smaller
constituent quark masses are more contributive to the
magnetic form factors.  However, as $Q^2$ increases,
the dependence on the constituent quark mass
undergoes a change, {\em i.e.} the greater
constituent quark masses contribute more to the
magnetic form factors.
In fact, we can reach the empirical data in the vicinity
of $Q^2=0$ with $M=370\mbox{MeV}$, we reproduce roughly
the correct momentum-dependence.  We select
$M=420\mbox{MeV}$ for the best fit to be consistent
with all observables in this paper.

Figures 10-11 present the contribution of the strange quark mass.
On the contrary to the electric form factors,
the $m_s$ correction enhances the magnetic form factors
about 5$\%$ to $10\%$.  In particular, it is of great significance
for the neutron magnetic form factor in fitting the
empirical data as shown in Figure 11.
 Our theoretical magnetic form factors
are in a good agreement with the empirical data within about
$15\%$ like the other quantities.

Table 2 shows each contribution of
the rotational $1/N_c$ and $m_s$
corrections to the magnetic moments, {\em i.e.}
$G^{B}_{M}(Q^2)$ at $Q^2=0$ (in Ref.\ \cite{kbpg}, the magnetic moments
are discussed in detail).  Our results are compared with the
SU(3) Skyrme model with pseudoscalar vector meson~\cite{pw}.
Figures 12-13 display the magnetic form factors of the charged
and neutral octet baryons, respectively.
The explicit breaking of $U$ spin symmetry in the magnetic form factors
are observed.
The corresponding magnetic charge radii are defined by
\begin{equation}
\langle r^2 \rangle^{B}_{M} \;=\; -\frac{6}{\mu_B} \left.
\frac{dG^{B}_{M} (Q^2)}{dQ^2} \right|_{Q^2=0}.
\label{Eq:magr}
\end{equation}
Their numerical results are listed in table 3.
The results for the nucleon are in a good agreement with the
experimental data.

\section{Summary and Conclusion}
The aim of this work has been to investigate the electromagnetic
form factors of the SU(3) octet baryons and related quantities
such as electromagnetic charge radii and magnetic moments
in the SU(3) semibosonized NJL model.
Starting from the effective chiral action, we have expressed
the matrix elements of electromagnetic current in the model.
When quantizing the soliton, the contributions arising from
the non-commutativity of collective operators was considered.
It gives a non-zero contribution of the rotational $1/N_c$ corrections.
The $m_s$ corrections are treated perturbatively,
the collective wave function correction being taken heed of.
The octet states of the baryon are mixed with higher
irreducible representations due to $m_s$.

The parameters of the model, including the cut-off, are adjusted
to $m_\pi=139\mbox{MeV}$ and $f_\pi=93\mbox{MeV}$.
The only parameter we have in the model is the constituent
quark mass $M$ which is fixed to $M=420\;\mbox{MeV}$
by the mass splitting of the SU(3) baryons.
The electric form factor of the proton is in an excellent
agreement with the empirical data.
As far as the electric form factor of the neutron is concerned,
it is well known that there are large uncertainties
in extracting it from experiments~\cite{edenetal}.
However, compared to ref.~\cite{meyer}, our result is found to be
in a remarkable agreement with it.
The electric charge radii of the nucleon
are also obtained in a good agreement with the experimental result
within about 10$\%$.

We also evaluated electric and magnetic form factors
of all other members of the SU(3) baryon octet.
The magnetic moments are in a good agreement with the experimental
data.  As far as the $Q^2$--dependence is concerned,
since there are no experimental data available,
these numbers are predictions.
 In all cases the $m_s$ corrections are about $10\%$.

Electromagnetic form factors of the baryons are used in order to
extract strange form factors from the experimental
data.  The evaluation of these quantities and of
semileptonic and mesonic decays of hyperons
will be the next steps in our research.

\section*{Acknowledgement}
We would like to thank Ch. Christov, P. Pobylitsa,
M. Prasza\l owicz, and T. Watabe
for fruitful discussions and critical comments.
This work has partly been supported by the BMFT, the DFG
and the COSY--Project (J\" ulich).  The work of M.P. is supported
in part by grant INTAS-93-0283.

\appendix
\section{The derivation of the regularization}
In this appendix, we shall give an explicit derivation of the
regularized $\Omega^0$ and $\Omega^1$ contributions
to the electromagnetic form factors.  We make use of the
proper-time regularization scheme.
We can see that the procedure
is very similar to the case of the axial constants~\cite{bpg2}.
Note that the non-anomalous part is regularized.
As is written in~eq.~(\ref{Eq:propr}),
the regularized effective action is expressed as
\begin{equation}
{\mbox Re} S_{eff} \;=\; {\rm Sp}\ \int \frac{du}{u}
\phi (u;\Lambda_i) \exp{(-uDD^\dagger)}  ,
\end{equation}
where
\begin{eqnarray}
D & = & \partial_\tau \;+\; H \; + \;i \Omega_E
\;+ \; \beta A^\dagger \hat{Q} A \;-\; i A_4 A^\dagger \hat{Q} A
\; - \; \alpha_k A_k A^\dagger \hat{Q} A  \nonumber \\
D^\dagger & = & -\partial_\tau \;+\; H \; - \;i \Omega_E
\;+ \; \beta A^\dagger \hat{Q} A \;+\; i A_4 A^\dagger \hat{Q} A
\; - \; \alpha_k A_k A^\dagger \hat{Q} A  .
\end{eqnarray}
Hence,
\begin{eqnarray}
DD^\dagger & = & W_0 (A^{0}_{\mu},\Omega^{0}, m^{0})
\;+\; W_1 (A^{1}_{\mu},\Omega^{0}, m^{0})
\;+\; W_2 (A^{1}_{\mu},\Omega^{1}) \nonumber \\
& + & W_3 (A^{0}_{\mu},\Omega^{1})
\;+\; W_4 (m^{1}) \;+\; O(\Omega^{1}, m^{1})
\;+\; O(\Omega^{2})\;+\; O(m^{2})
\end{eqnarray}
with
\begin{eqnarray}
W_0 & = & -\partial^{2}_{\tau} \;+\; H^{2}_{E} \nonumber \\
W_1 & = & i\{A_4 A^\dagger \hat{Q} A, \;\partial_\tau\}
\;-\; [\alpha_k A_k A^\dagger \hat{Q} A,\; \partial_\tau]
\nonumber \\
& - & i [H_E, \;A_4 A^\dagger \hat{Q} A]
\;-\; \{H_E, \;\alpha_k A_k A^\dagger \hat{Q} A\}\nonumber \\
W_2 & = & -\{\Omega_E, \;A_4 A^\dagger \hat{Q} A\}
\;+\;i [\Omega_E, \;\alpha_k A_k A^\dagger \hat{Q} A]
\nonumber \\
W_3 & = & -i \{\Omega_E,\;\partial_\tau\}
\;+\;i[H_E,\;\Omega_E]
\nonumber \\
W_4 & = & [\beta A^\dagger \hat{m} A, \;\partial_\tau]
\;+\;\{H_E,\;\beta A^\dagger \hat{m} A \}
\nonumber \\
& - & i[\beta A^\dagger \hat{m} A, \;A_4 A^\dagger \hat{Q} A]
\;+\; \{\beta A^\dagger \hat{m} A,
\;\alpha_k A_k A^\dagger \hat{Q} A\}.
\end{eqnarray}
The terms of higher orders in $\Omega$ and $\hat{m}$ and
of $\Omega \cdot \hat{m}$ are neglected, since they are
believed to be very tiny.

Taking advantage of the Feynman-Schwinger-Dyson formula,
we can expand $\exp{(-uW)}$ around $W_0$:
\begin{eqnarray}
\exp{(-uW)} & = & \exp{(-uW_0)} \nonumber \\
& - & u \int^{1}_{0} d\alpha \exp{(-u\alpha W_0)}
[W \;-\;W_0] \exp{(-u(1-\alpha)W_0)} \nonumber \\
& + & u^2 \int^{1}_{0} d\beta \int^{1-\beta}_{0}
\exp{(-u\alpha W_0)}[W\;-\;W_0] \times \nonumber \\
&\times & \exp{(-u\beta W_0)}
[W\;-\;W_0]\exp{(-u(1-\alpha-\beta)W_0)} \nonumber \\
& + & \;\cdots
\end{eqnarray}
First, we shall consider in case of the electric form factor.
The lowest order contribution of $\Omega_E$ vanishes.
The sea contribution of $\Omega^{0}_{E}$ comes only
from the imaginary part of the effective action.
As for the next order of $\Omega_E$,
we need $W_2$ and $W_1 \cdot W_3$.
After some manipulations, we obtain
\begin{eqnarray}
\langle B,p' |V_{0} (0) | B,p \rangle^{\Omega^{1}}  & = &
\frac{N_c}{4} \sum_{nm} R_I (E_n, E_m)
\langle \{D^{(8)}_{Qa},\;i\Omega^{b}_{E} \} \rangle_B
\times \nonumber  \\
& \times & \int d^3 x e^{i\vec{q}\cdot\vec{x}} \int d^3y
\Psi^{\dagger}_{n} (x) \lambda^a \Psi_{m}(x)
\Psi^{\dagger}_{m} (y) \lambda^b \Psi_{n}(y) .
\end{eqnarray}
The $m_s$ correction due to $W_4$ and $W_4 \cdot W_1$
vanishes like $\Omega^0$ contribution.
The $m_s$ correction arises only from
the quantization of $i\Omega_E$~\cite{betal2}.

The regularization of the magnetic form factor is
more involved due to the time-ordering of collective
operators.
Here, we need only the term
$-A_k \{H_E,\;\alpha_k A^\dagger \hat{Q} A\}$ for the
lowest order contribution:
\begin{eqnarray}
\langle B,p' | V_i(0) | B,p \rangle^{\Omega^{0}}
& = & \frac{1}{2}
\frac{\delta}{\delta A_i} {\mbox{Sp}}
\int du \phi (u;\Lambda_i) \int^{1}_{0} d\alpha
\exp{(-u\alpha W_0)} \nonumber \\
&\times &
A_k \{H_E,\;\alpha_k A^\dagger \hat{Q} A\}
\exp{(-u(1-\alpha)W_0)} \nonumber \\
& = & \frac{N_c}{2} D^{(8)}_{Qa} \sum_{n}
\langle n | \alpha_i  \lambda^a
| n \rangle
R(E_n),
\end{eqnarray}
where $R(E_n)$ is defined in~eq.~(\ref{Eq:regul}).

As a next step, we proceed to evaluate the $\Omega^{1}_{E}$
correction to the magnetic form factor.  It is
tedious but straightforward:
\begin{equation}
\langle B,p' | V_i(0) | B,p\rangle^{\Omega^{1}}\;=\;
\frac{\delta}{\delta A_i} \left (X_1[A_k] + X_2 [A_k] \right)
\left.\right|_{A_k = 0},
\end{equation}
where
\begin{eqnarray}
X_1[A_k] & = &  \frac{1}{2}   {\mbox{Sp}}
\int du \phi (u;\Lambda_i) \int^{1}_{0} d\alpha
\exp{(-u\alpha W_0)} \nonumber \\
&\times & W_2[A_k]
\exp{(-u(1-\alpha)W_0)} , \\
X_2[A_k] & = & \frac{1}{2}   {\mbox{Sp}}
\int du \phi (u;\Lambda_i) \int^{1}_{0} d\beta
\int^{1-\beta}_{0} d\alpha
\exp{(-u\alpha W_0)}\nonumber \\
&\times &
\left( W_1[A_k] + W_3[A_k]\right)\exp{(-u\beta W_0)}
\left( W_1[A_k] + W_3[A_k]\right) \nonumber \\
& & \exp{(-u(1-\alpha - \beta) W_0)}.
\end{eqnarray}
The terms including $W_1 \cdot W_1$ and $W_3\cdot W_3$ vanish.
The first term $\frac{\delta}{\delta A_i}X_1[A_k]$
is obtained to be
\begin{eqnarray}
\frac{\delta}{\delta A_i}X_1[A_k] & = & \frac{-i}{16}
N_c \sum_{n,m} \sqrt{\frac{u}{\pi}}
(e^{-uE^{2}_{n}} - e^{-uE^{2}_{m}})
\{ i\Omega^{a}_{E}, D^{(8)}_{Qb} \}\nonumber \\
& & \langle n | \lambda^{a} | m \rangle
\langle m | \alpha_i \lambda^{b} | m \rangle .
\end{eqnarray}
The second term is as follows:
\begin{eqnarray}
\frac{\delta}{\delta A_i}X_2[A_k] & = &
-u\frac{iN_c}{8\pi}\sum_{n,m} \int^{1}_{0}d\beta
e^{-u[\beta E^{2}_{m} + (1-\beta)E^{2}_{n}]} \nonumber \\
& & \left[\beta E_{m} - (1-\beta)E_{n}\right]
\frac{1}{\sqrt{\beta (1-\beta)}}
\left[i\Omega^{a}_{E}, D_{Qb}\right] \nonumber \\
& & \langle n | \lambda^{a} | m \rangle
\langle m | \alpha_i \lambda^{b} | n \rangle
\nonumber \\
& + & \frac{i}{16}
N_c \sum_{n,m} \sqrt{\frac{u}{\pi}}
(e^{-uE^{2}_{n}} - e^{-uE^{2}_{m}})
\{ i\Omega^{a}_{E}, D^{(8)}_{Qb} \}\nonumber \\
& & \langle n | \lambda^{a} | m \rangle
\langle m | \alpha_i \lambda^{b} | n \rangle .
\label{Eq:second}
\end{eqnarray}
The second part of eq.~(\ref{Eq:second}) is cancelled by
$\frac{\delta}{\delta A_i}X_1[A_k]$, so that we have
\begin{eqnarray}
\langle B,p' | V_i(0) | B,p \rangle^{\Omega^{1}} & = &
-u\frac{iN_c}{8\pi}\sum_{n,m} \int^{1}_{0}d\beta
e^{-u[\beta E^{2}_{m} + (1-\beta)E^{2}_{n}]} \nonumber \\
& & \left[\beta E_{m} - (1-\beta)E_{n}\right]
\frac{1}{\sqrt{\beta (1-\beta)}}
\left[i\Omega^{a}_{E}, D_{Qb}\right] \nonumber \\
& & \langle n | \lambda^{a} | m \rangle
\langle m | \alpha_i \lambda^{b} | n \rangle .
\end{eqnarray}
Having integrated over $\beta$, we obtain
\begin{equation}
\langle B,p'| V_i(0) |B,p \rangle^{\Omega^{1}} \;= \;
-\frac{N_c}{4} \sum_m
\langle [D^{(8)}_{Qa}, J_b]\rangle_B 
\langle n | \lambda^{a} | m \rangle
\langle m | \alpha_i \lambda^{b} | n \rangle
{\cal R}_{\cal Q} (E_n,E_m),
\end{equation}
where ${\cal R}_Q$ is defined in~eq.~(\ref{Eq:regul}).

\section{The gradient expansion of the magnetic moments}
It is well known that the exact expressions for the magnetic moments
can be expanded in powers of gradients of the chiral fields
\cite{ait}. In this way the quark determinant gives terms, which are
quite similar to the Skyrme model expressions \cite{pw}.
An important difference is however the contributions of order
$\Omega^1$ from the real part of the action.
In the present case we obtain
\begin{eqnarray}
\mu_B  &=&   - 2 M_n \int dr\  r^2 \sin^2\theta
      \langle D_{Q3}  \rangle_B
     \left[ {8\pi\over 3} f_\pi^2 + {1\over 3}
       {M_{u}\over 4I_1}  +  {1\over 3} {M_{u}\over
         8I_2}
     \right]      \nonumber \\  & &
     +{4 \over 9\pi}   \int dr r^2 \sin^2\theta\  \theta' \
      \left[  { -\langle d_{3pp} D_{Qp} J_p \rangle_B \over I_2}
    - { \langle D_{Q8} J_3 \rangle_B \over I_1 \sqrt{3} }  \right].
\end{eqnarray}
Our numerical densities for the electromagnetic form factors
are compared with those obtained from the gradient expansion
in order to warrant the calculation.

\vfill\eject
\begin{table}[]
\caption{The electric charge radii of the SU(3) octet baryons
predicted by our model compared to the evaluation
from the Skyrme model by Park and Weigel [37] and the experimental
numbers.  The constituent quark mass is fixed to $M=420$ MeV.}
\begin{tabular}{cccc}
$\mbox{Baryons}$   & Our model & Park $\&$ Weigel &
Experiment \\
\hline
$p$ & $\phantom{-}0.78$ & $\phantom{-}1.20$ &$\phantom{-}0.74$ \\
$n$ & $-0.09$ & $-0.15$ & $-0.11$ \\
$\Lambda$ & $-0.04$ & $-0.06$ & -- \\
$\Sigma^{+}$ & $\phantom{-}0.79$ & $\phantom{-}1.20$ & -- \\
$\Sigma^{0}$ & $\phantom{-}0.02$ & $-0.01$ & -- \\
$\Sigma^{-}$ & $-0.75$ & $-1.21$ & -- \\
$\Xi^{0}$ & $-0.06$ & $-0.10$ & --\\
$\Xi^{-}$ & $-0.72$ & $-1.21$ & -- \\
\end{tabular}
\end{table}
\begin{table}[]
\caption{The magnetic moments of the SU(3) octet baryons
predicted by our model.  Each contribution is listed
from the leading order.  The results are also compared
with the Skyrme model of Park and Weigel [37].
The experimental data for the magnetic moments
are taken from Ref.[45].  Our final values are given
by $\mu_B (\Omega^1, m^{1}_{s})$.
The constituent quark mass is fixed to $M=420$ MeV.}
\begin{tabular}{cccccc}
$\mbox{Baryons}$   &
$\mu_B (\Omega^0, m^{0}_{s})$ &
$\mu_B(\Omega^1, m^{0}_{s})$ &
$\mu_B(\Omega^1, m^{1}_{s})$ &
Park $\&$ Weigel &
$\mbox{Exp.}$ \\
\hline
$p$ & $\phantom{-}1.01$ & $\phantom{-}2.27$
&$\phantom{-}2.39$ & $\phantom{-}2.36$ & $\phantom{-}2.79$ \\
$n$ & $-0.75$ & $-1.55$ & $-1.76$ & $-1.87$ & $-1.91$ \\
$\Lambda$ & $-0.38$ & $-0.78$ & $-0.77$ & $-0.60$ & $-0.61$ \\
$\Sigma^{+}$ & $\phantom{-}1.01$ & $\phantom{-}2.27$ &
$\phantom{-}2.42$ & $\phantom{-}2.41$ & $\phantom{-}2.46$  \\
$\Sigma^{0}$ & $\phantom{-}0.38$ & $\phantom{-}0.78$
& $\phantom{-}0.75$ & $\phantom{-}0.66$ &-- \\
$\Sigma^{-}$ & $-0.25$ & $-0.71$ & $-0.92$ & $-1.10$ & $-1.16$ \\
$\Xi^{0}$ & $-0.75$ & $-1.55$ & $-1.64$ & $-1.96$ & $-1.25$ \\
$\Xi^{-}$ & $-0.25$ & $-0.71$ & $-0.68$ & $-0.84$ & $-0.65$ \\
$|\Sigma^0\rightarrow \Lambda|$ & $\phantom{-}0.65$
&$\phantom{-}1.34$ &$\phantom{-}1.51$ &$\phantom{-}1.74$
&$\phantom{-}1.61$ \\
\end{tabular}
\end{table}
\begin{table}[]
\caption{The magnetic charge radii of the SU(3) octet baryons
predicted by our model compared with the Skyrme model of
Park and Weigel [37].
The constituent quark mass is fixed to $M=420$ MeV.}
\begin{tabular}{cccc}
$\mbox{Baryons}$   & Our model & Park $\&$ Weigel &
Experiment \\
\hline
$p$ & $0.70$ & $0.94$ &$0.74$ \\
$n$ & $0.78$ & $0.94$ & $0.77$ \\
$\Lambda$ & $0.70$ & $0.78$ & -- \\
$\Sigma^{+}$ & $0.71$ & $0.96$ & -- \\
$\Sigma^{0}$ & $0.70$ & $0.86$ & -- \\
$\Sigma^{-}$ & $0.74$ & $1.07$ & -- \\
$\Xi^{0}$ & $0.75$ & $0.90$ & --\\
$\Xi^{-}$ & $0.51$ & $0.84$ & -- \\
\end{tabular}
\end{table}

\vfill\eject
\newpage
\begin{center}
{\Large Figure Captions}
\end{center}

{\bf Fig.~1}:
 The proton electric formfactor as a function of $Q^2$:
The dashed curve corresponds to
the constituent quark mass $M=370 \mbox{MeV}$, while
solid curve is for $M=420\mbox{MeV}$.  The dotted curve
displays the case of $M=450\mbox{MeV}$.
The empirical data are taken from
H\" ohler {\em et al.}~\cite{holetal}.
\vspace{0.5cm}

{\bf Fig.~2}:
 The neutron electric formfactor as a function of $Q^2$:
The solid curve corresponds to
the constituent quark mass M=420 MeV, while
dashed curve draws M=370 MeV.  The dotted curve
displays the case of M=450 MeV.
The empirical data are taken from Platchkov {\em et al.}~\cite{pl}.
The other four points are results for $G^{n}_{E}$
extracted by Woodward {\em et al.}~\cite{elecn1} (open diamond),
by Thompson {\em et al.}~\cite{elecn2} (open box),
by Eden {\em et al.}~\cite{edenetal} (open circle)
and by Meyerhoff {\em et al.}~\cite{meyer} (open triangle).
\vspace{0.5cm}

{\bf Fig.~3}:
 The proton electric formfactor as a function of $Q^2$:
The solid curve corresponds to
the strange quark mass $m_s=180\; \mbox{MeV}$, while
dashed curve draws without $m_s$.  The dotted curve
displays the case of the SU(2) model.
$M=420 \;\mbox{MeV}$ is chosen for the constituent quark mass.
The empirical data are taken from
H\" ohler {\em et al.}~\cite{holetal}.
\vspace{0.5cm}

{\bf Fig.~4}:
 The neutron electric formfactor as a function of $Q^2$:
The solid curve corresponds to
the strange quark mass $m_s=180 \;\mbox{MeV}$, while
dashed curve draws without $m_s$.  The dotted curve
displays the case of the SU(2) model.
$M=420 \; \mbox{MeV}$ is chosen for the constituent quark mass.
The empirical data(shaded circle) are taken from
Platchkov {\em et al.}~\cite{pl}.
The other four points are results for $G^{n}_{E}$
extracted by Woodward {\em et al.}~\cite{elecn1} (open diamond),
by Thompson {\em et al.}~\cite{elecn2} (open box),
by Eden {\em et al.}~\cite{edenetal} (open circle) and
by Meyerhoff {\em et al.}~\cite{meyer} (open triangle).
\vspace{0.5cm}

{\bf Fig.~5}:
 The electric isospin formfactors of the nucleon as a function of $Q^2$:
The solid curve corresponds to the isoscalar electric formfactor
of the nucleon in SU(3), while the dashed curve denotes the isovector
one.  The dot-dashed curve draws the isoscalar one in SU(2),
whereas the dotted curve stands for the isovector one in SU(2).
\vspace{0.5cm}

{\bf Fig.~6}:
 The electric formfactors of the charged SU(3) octet baryons
as a function of $Q^2$:
The solid curve corresponds to the proton electric formfactor.
The dashed curve is for $\Sigma^{+}$.  The dash-dotted curve
displays that of $\Sigma^{-}$. The dotted curve
represents that of $\Xi^{-}$.
$M=420 \; \mbox{MeV}$ is chosen for the constituent quark mass.
\vspace{0.5cm}

{\bf Fig.~7}:
 The electric formfactors of the neutral SU(3) octet baryons
as a function of $Q^2$:
The solid curve corresponds to the neutron electric formfactor.
The dashed curve is for $\Lambda$.  The dash-dotted curve
displays that of $\Sigma^{0}$. The dotted curve
represents that of $\Xi^{0}$.
$M=420 \; \mbox{MeV}$ is chosen for the constituent quark mass.
\vspace{0.5cm}

{\bf Fig.~8}:
 The proton magnetic formfactor as a function of $Q^2$:
The dashed curve corresponds to
the constituent quark mass $M=370\; \mbox{MeV}$, while
solid curve is for $M=420\;\mbox{MeV}$.  The dotted curve
displays the case of $M=450 \;\mbox{MeV}$.
The empirical data are taken from
H\" ohler {\em et al.}~\cite{holetal}.
The numbers are given in units of the Bohr-magneton
without any rescaling.
\vspace{0.5cm}

{\bf Fig.~9}:
 The neutron magnetic formfactor as a function of $Q^2$:
The dashed curve corresponds to
the constituent quark mass $M=370\; \mbox{MeV}$, while
solid curve is for $M=420 \;\mbox{MeV}$.  The dotted curve
displays the case of $M=450 \;\mbox{MeV}$.
The empirical data represented by black dots
are taken from H\" ohler {\em et al.}~\cite{holetal} while
the data with open triangles are due to the most recent
experiment~\cite{Bruinsetal}.
The numbers are given in units of the Bohr-magneton
without any rescaling.
\vspace{0.5cm}

{\bf Fig.~10}:
 The proton magnetic formfactor as a function of $Q^2$:
The solid curve corresponds to
the strange quark mass $m_s=180 \mbox{MeV}$, while
dashed curve draws without $m_s$.  The dotted curve
displays that of the SU(2) model.
$M=420 \;\mbox{MeV}$ is chosen for the constituent quark mass.
The empirical data are taken from
H\" ohler {\em et al.}~\cite{holetal}.
The numbers are given in units of the Bohr-magneton
without any rescaling.
\vspace{0.5cm}

{\bf Fig.~11}:
 The neutron magnetic formfactor as a function of $Q^2$:
The solid curve corresponds to
the strange quark mass $m_s=180\; \mbox{MeV}$, while
dashed curve draws without $m_s$.  The dotted curve
displays the case of the SU(2) model.
$M=420 \;\mbox{MeV}$ is chosen for the constituent quark mass.
The empirical data represented by black dots
are taken from H\" ohler {\em et al.}~\cite{holetal} while
the data with open triangles are due to the most recent
experiment~\cite{Bruinsetal}.
The numbers are given in units of the Bohr-magneton
without any rescaling.
\vspace{0.5cm}

{\bf Fig.~12}:
 The magnetic formfactors of the charged SU(3) octet baryons
as a function of $Q^2$:
The solid curve corresponds to the proton magnetic formfactor.
The dashed curve is for $\Sigma^{+}$.  The dash-dotted curve
displays that of $\Sigma^{-}$. The dotted curve
represents that of $\Xi^{-}$.
The experimental data for the magnetic moments are taken from
Ref~\cite{pdgroup}.
$M=420 \;\mbox{MeV}$ is chosen for the constituent quark mass.
\vspace{0.5cm}

{\bf Fig.~13}:
 The magnetic formfactors of the neutral SU(3) octet baryons
as a function of $Q^2$:
The solid curve corresponds to the neutron magnetic formfactor.
The dashed curve is for $\Lambda$.  The dash-dotted curve
displays that of $\Sigma^{0}$. The dotted curve
represents that of $\Xi^{0}$.
The experimental data for the magnetic moments are taken from
Ref.~\cite{pdgroup}.
$M=420\; \mbox{MeV}$ is chosen for the constituent quark mass.

\pagebreak
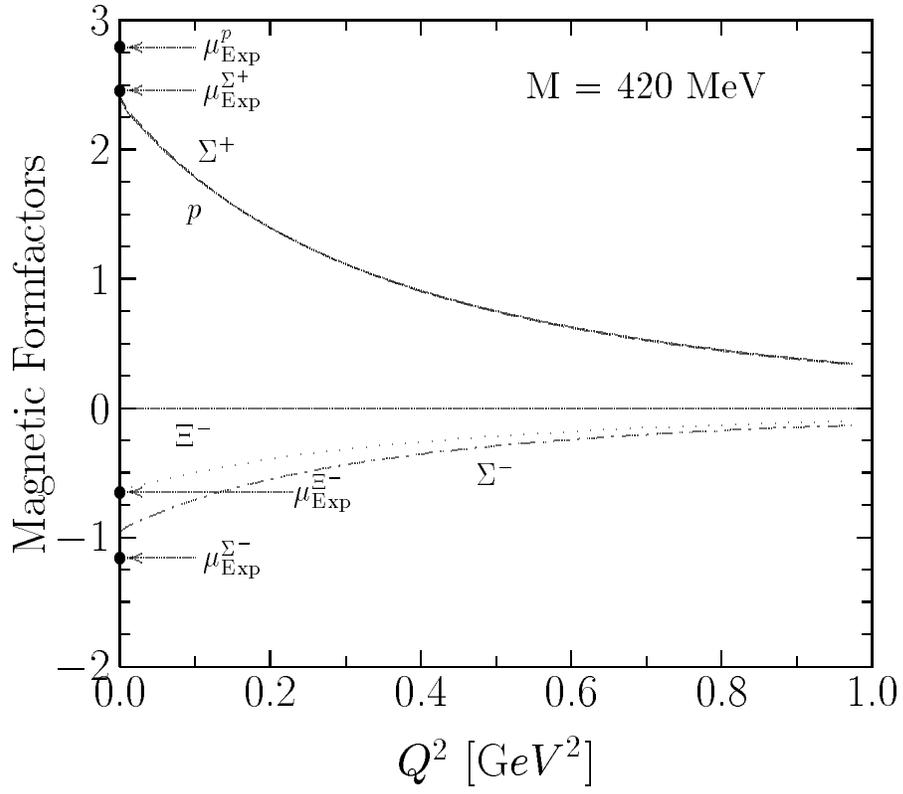
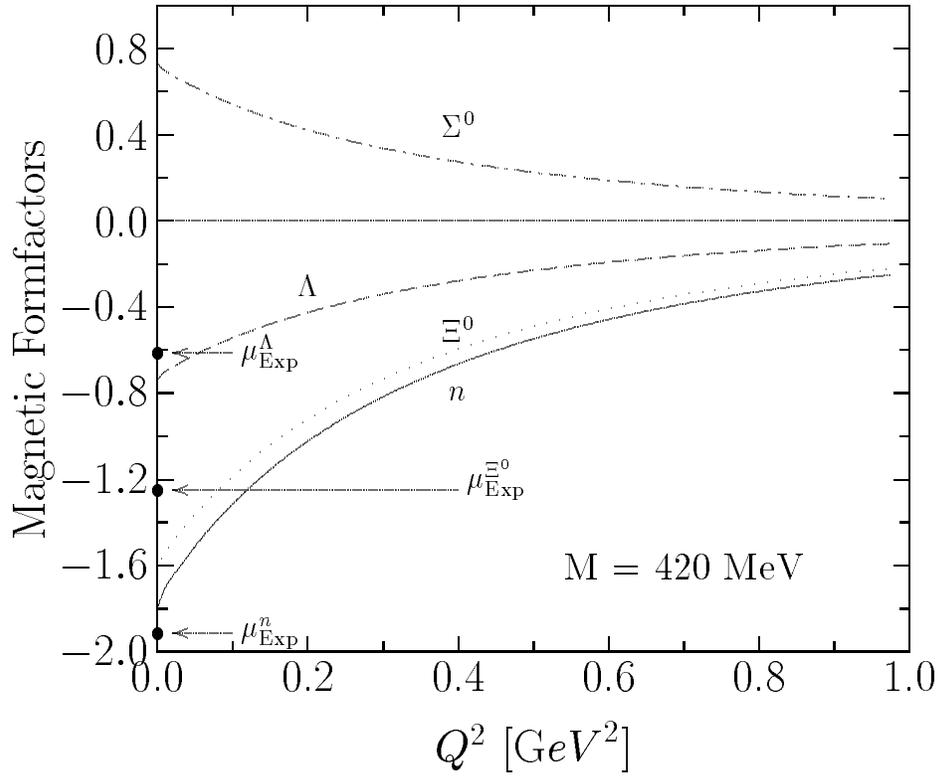
\end{document}